\documentclass[twocolumn,showpacs,showkeys,prd,aps,amsmath,amssymb,10pt]{revtex4-1}
\usepackage{graphicx}
\usepackage{color}
\usepackage{multirow}
\usepackage{bm}
\usepackage{mathtools}
\usepackage[caption=false]{subfig}
\usepackage{url}
\usepackage{breakurl}
\usepackage{physics}
\usepackage{cancel}
\allowdisplaybreaks

\begin{document}

\title{Towards an effective field theory approach to the neutrinoless double-beta decay}

\author{Mihai Horoi}
\email{mihai.horoi@cmich.edu}
\author{Andrei Neacsu}
\email{neacs1a@cmich.edu}
\affiliation{Department of Physics, Central Michigan University, Mount Pleasant, Michigan 48859, USA}
\date{\today}
\begin{abstract}
Weak interaction in nuclei represents a well-known venue for testing many of the fundamental symmetries of the Standard Model. 
In particular, neutrinoless double-beta decay offers the possibility to test Beyond Standard Model theories predicting that neutrinos are Majorana fermions and the lepton number conservation is violated.
This paper focuses on an effective field theory approach to neutrinoless double-beta decay for extracting information regarding the properties of the Beyond Standard Model Lagrangian responsible for this process. 
We use shell model nuclear matrix elements and the latest experimental lower limits for the half-lives to extract the lepton number violating parameters of five nuclei of experimental interest, and lower limits for the energy scales of the new physics.
\end{abstract}

\maketitle

\newcommand*{\vv}[1]{\vec{\mkern0mu#1}}

\section{Introduction}
The neutrinoless double-beta decay ($0\nu\beta\beta$) is considered the best approach to study the yet unknown properties of neutrinos related to their nature, whether they are Dirac or Majorana fermions, which the neutrino oscillation experiments cannot clarify. Should the neutrinoless double-beta transitions occur, then the lepton number conservation is violated by two units, and the black-box theorems~\cite{SchechterValle1982,Nieves1984,Takasugi1984,Hirsch2006} indicate that the light left-handed neutrinos are Majorana fermions.  As such, through black-box theorems alone, it is not possible to disentangle the dominant mechanism contributing to this process. 
Most of the theoretical effort dedicated to this subject consists of calculations of leptonic phase-space factors and nuclear matrix elements that are computed via several nuclear structure methods and within specific models. One of the most popular models is the left-right symmetric model 
\cite{PatiSalam1974,MohapatraPati1975,Senjanovic1975,KeungSenjanovic1983,Barry2013}, which is currently investigated at the Large Hadron Collider (LHC)~\cite{CMS2014}.
In two recent papers~\cite{HoroiNeacsu2016prd,Neacsu2016ahep-dist} we have discussed ways to identify some of the possible contributions to the decay rate by studying the angular distribution and the energy distribution of the two outgoing electrons that could be measured. However, there are still many other possible contributions to this process that one cannot yet dismiss. 
For these reasons, a more general beyond standard model (BSM) effective field theory would be preferable, as it would not be limited to relying on specific models, but rather considering the most general BSM effective field theoretical approach that describes this process. An important outcome of such a theory is the evaluation of the energy scales up to which the effective field operators are not broken, together with limits for the effective low-energy couplings. 

The analysis of the $0\nu\beta\beta$ decay process is generally done at three levels. At the lowest level the weak interaction of the quarks and leptons is considered and the BSM physics is treated within a low-energy effective field theory approach. At the next level the hadronization process to nucleons and exchanging pion is considered. The nucleons are treated in the impulse approximation leading to free space $0\nu\beta\beta$ transition operators. At the third level the nucleon dynamics inside the nuclei is treated using nonperturbative nuclear wave functions, which are further used to obtain nuclear matrix elements (NME) needed to calculate the $0\nu\beta\beta$ observables, such as half-lives and two-electron angular and energy distributions~\cite{HoroiNeacsu2016prd}. A modern approach that could accomplish this plan would based on the chiral effective field theory of pions and nucleons \cite{Cirigliano2017-arxiv,Cirigliano2017-PLB}. This approach introduces a number of couplings, which in principle can be calculated from the underlying theory of strong interaction using lattice QCD techniques \cite{Cirigliano2017-arxiv}, or may be extracted within some approximation from the known experimental data \cite{Cirigliano2017-PLB}. These couplings may come with new phases and they may include effective contributions from the exchange of heavier mesons. The lattice QCD approach is underway, but it proved to be very difficult for extracting even basic weak nucleon couplings, such as $g_A$ \cite{Loud2017}.

In this paper we start from the formalism of Ref.~\cite{Hirsch1996plb,Pas1999,Pas2001,Deppisch2012} that provides a general effective field theory (EFT) approach to the neutrinoless double-beta decay. However, at the hadron level three new diagrams are added for the first time to the effective field theory analysis of the $0\nu\beta\beta$, which were only considered in the literature in the context of specific mechanisms. 
Under the assumption that a single coupling in the BSM Lagrangian dominates the $0\nu\beta\beta$ amplitude, we extract new limits for the effective Majorana mass and for 11 additional low-energy EFT couplings using data from five nuclei of current experimental interest. Some of these couplings correspond to parameters found in left-right symmetric models, and we present and compare them.
To be able to get the limits of these effective couplings and parameters from the experimental half-life limits, 23 nuclear matrix elements (NME) and 9 phase-space factors (PSF) are needed. Finally, we use the limits for the EFT couplings and the formalism of the effective field theory to obtain limits for the energy scale of the new physics that could be responsible for the neutrinoless double beta decay process.

To accomplish this goal we need reliable NME. The most commonly used nuclear structure methods for the NME calculation are 
proton-neutron Quasi Random Phase Approximation (pnQRPA)~\cite{Hirsch1996plb,Pas1999,Pas2001,Deppisch2012,Simkovic1999,Suhonen2010, Faessler2011, MustonendEngel2013, FaesslerGonzales2014}, 
Interacting Shell Model (ISM)~\cite{Retamosa1995,Caurier1996, Horoi2013,Neacsu2014constraints, Caurier2008, MenendezPovesCaurier2009, Caurier2005, HoroiStoica2010,NeacsuStoicaHoroi2012, SenkovHoroi2013, HoroiBrown2013, SenkovHoroiBrown2014,BrownHoroiSenkov2014, NeacsuStoica2014, SenkovHoroi2014, NeacsuHoroi2015,NeacsuHoroi2016,HoroiStoicaBrown2007,Blennow2010}, 
Interacting Boson Model (IBM-2)~\cite{Barea2009, Barea2012, Barea2013, Barea2015}, Projected Hartree Fock Bogoliubov (PHFB)~\cite{Rath2013}, Energy Density Functional (EDF)~\cite{Rodriguez2010}, and the Relativistic 
Energy Density Functional (REDF)~\cite{Song2014} method. The NME calculated with different methods and by different groups sometimes show large differences, and this has been debated in the literature~\cite{Faessler2012,Vogel2012}. Although there seem to exist many NME results to choose from, most of the references listed only provide calculations for the light left-handed Majorana neutrino exchange. 
Ref.~\cite{NeacsuHoroi2016} provides tables and plots that compare the latest results for the light left-handed neutrino exchange and for the heavy right-handed neutrino exchange. 

The NME used in Ref.~\cite{Hirsch1996plb,Pas1999,Pas2001,Deppisch2012} come from older QRPA calculations, which do not include many of the improvements proposed in recent years~\cite{SimkovicRodin2013,Suhonen2015}. 
We calculate the NME using shell model techniques, which are consistent with previous calculations~\cite{HoroiStoicaBrown2007,HoroiStoica2010,NeacsuStoicaHoroi2012,
HoroiBrown2013,NeacsuStoica2014,SenkovHoroi2013,Horoi2013,BrownHoroiSenkov2014,SenkovHoroi2014,SenkovHoroiBrown2014,NeacsuHoroi2015,NeacsuHoroi2016}. The reason for choosing shell model NME is our belief that these are better suited and more reliable for $0\nu\beta\beta$ calculations, as they take into account all the correlations around the Fermi surface, respect all symmetries, and take into account consistently the effects of the missing single particle space via many-body perturbation theory (the effects were shown to be small, about 20\%, for $^{82}$Se~\cite{HoltEngel2013}). 
Furthermore, we have tested the shell model methods and the effective Hamiltonians used by comparing calculations of spectroscopic observables to the experimental data, as presented in Ref.~\cite{HoroiStoica2010, Senkov2016, NeacsuHoroi2016}.
We do not consider any quenching for the bare $0\nu\beta\beta$ operator in these calculations. Such a choice is different from that for the simple Gamow-Teller operator used in the single beta and $2\nu\beta\beta$ decays where a quenching factor of about 0.7 is necessary~\cite{BrownFangHoroi2015}. 
For the PSF we use an effective theory based on the formalism of Ref.~\cite{Doi1985}, but fine-tuned as to take into account the effects of a Coulomb field distorting finite-size proton distribution in the final nucleus. To our knowledge, some of the NME presented in this paper are calculated for the first time using shell model techniques.

This paper is organized as follows: Section~\ref{section-mechanisms} analyzes the contributions of several BSM mechanisms to the neutrinoless double-beta decay.
Section~\ref{section-eft} presents the framework of the effective field theory for the neutrinoless double-beta decay.
Section~\ref{results} shows the experimental limits on the BSM lepton number violating (LNV) couplings that we calculate, and is divided into three subsections. 
Subsection~\ref{subsection-mass} is dedicated to the revisit of the most common approach to $0\nu\beta\beta$ that considers only
the light left-handed Majorana neutrino exchange, presenting shell model nuclear matrix elements and upper limits for the Majorana mass. 
Subsection~\ref{subsection-long} details the study of the long-range contributions to the $0\nu\beta\beta$ decay Lagrangian.
Subsection~\ref{subsection-short} presents the analysis of the short-range contribution LNV parameters. 
Discussions are presented in Section~\ref{discussions}, Section~\ref{section-conclusions} is dedicated to conclusions and, last, 
Section~\ref{section-appendix} is an Appendix containing all the relevant formulae for calculating the NME.

 \begin{table*}
  \caption{The $Q^{0\nu}_{\beta\beta}$ values in MeV, the experimental $T^{0\nu}_{1/2}$ limits in years, and the calculated PSF ($G_{01} - G_{09}$) in years$^{-1}$ for all five isotopes currently under investigation.}
  \begin{tabular}{lccccc} \label{tab-psf}
			   &$^{48}$Ca	&$^{76}$Ge	&$^{82}$Se	&$^{130}$Te	&$^{136}$Xe \\ \hline
$Q^{0\nu}_{\beta\beta}$~\cite{StoicaMirea2013}&4.272	&2.039		&2.995		&2.813		&2.287 \\ 		   
$T^{0\nu}_{1/2}\ >$	   &$2.0 \cdot 10^{22}$~\cite{Nemo3-2016} 
			   &$5.3 \cdot 10^{25}$~\cite{Gerda-2017} 
			   &$2.5 \cdot 10^{23}$~\cite{Waters-2016} 
			   &$4.0 \cdot 10^{24}$~\cite{Cuore-2015} 
			   &$1.07 \cdot 10^{26}$~\cite{kamlandzen16} 
			   \\ \hline
$G_{01}\cdot 10^{14}$	   &$2.45$ 	&$0.22$ 	&$1.00$		&$1.41$		&$1.45$ \\
$G_{02}\cdot 10^{14}$	   &$15.4$	&$0.35$		&$3.21$		&$3.24$		&$3.15$ \\
$G_{03}\cdot 10^{15}$	   &$18.2$	&$1.20$		&$6.50$		&$8.46$		&$8.55$ \\
$G_{04}\cdot 10^{15}$	   &$5.04$	&$0.42$		&$1.92$		&$2.53$		&$2.58$ \\
$G_{05}\cdot 10^{13}$	   &$3.28$ 	&$0.60$ 	&$2.16$		&$4.12$		&$4.36$ \\
$G_{06}\cdot 10^{12}$	   &$3.87$	&$0.50$		&$1.65$		&$2.16$		&$2.21$ \\
$G_{07}\cdot 10^{10}$	   &$2.85$	&$0.28$ 	&$1.20$		&$1.75$		&$1.80$ \\
$G_{08}\cdot 10^{11}$	   &$1.31$ 	&$0.17$ 	&$0.82$		&$1.72$		&$1.83$ \\
$G_{09}\cdot 10^{10}$	   &$15.5$	&$1.12$		&$4.42$		&$4.47$		&$4.44$ 
  \end{tabular}
 \end{table*}

\section{BSM mechanisms contributing to neutrinoless double-beta decay} \label{section-mechanisms}

The main mechanism considered to be responsible for the neutrinoless double beta decay is the mass mechanism that assumes that the neutrinos are Majorana fermions, and relies on the assumption that the light left-handed neutrinos have mass.
However, the possibility that right-handed currents could contribute to the neutrinoless double-beta decay $( 0\nu\beta\beta  )$ has been already considered for some time~\cite{Doi1983,Doi1985}. 
Recently, $0\nu\beta\beta$ studies~\cite{Rodejohann2012,Barry2013} have adopted the left-right symmetric model~\cite{Mohapatra1975,Senjanovic1975} for the inclusion of right-handed currents. In addition, the $R$-parity violating $(\cancel{\cal{R}}_p)$ supersymmetric (SUSY) model  can also contribute to the neutrinoless double beta decay process~\cite{Hirsch1996,Hirsch1997,Kovalenko2008}. 
In the framework of the left-right symmetric model and $R$-parity violating SUSY model, the $0\nu\beta\beta$ half-life can be written as a sum of products 
of PSF, BSM LNV parameters, and their corresponding NME~\cite{HoroiNeacsu2016prd}:
\begin{flalign}\label{lrsm-hl}
 \nonumber \left[ T^{0\nu}_{1/2} \right] ^{-1}=& \ G_{01} g^4_A  \left| \eta_{0\nu}M_{0\nu} +\left(\eta^L_{N_R}+\eta^R_{N_R}\right)M_{0N} \right. \\
 + & \left. \eta_{\tilde{q}}M_{\tilde{q}} +\eta_{\lambda'}M_{\lambda'} +\eta_{\lambda}X_{\lambda}+\eta_{\eta}X_{\eta} \right| ^2  .
\end{flalign}
Here, $G_{01}$ is a phase-space factor that can be calculated with good precision for most cases~\cite{SuhonenCivitarese1998,Kotila2012,StoicaMirea2013,HoroiNeacsu2016psf},
$g_A$ is the axial vector coupling constant, $\eta_{0\nu}=\frac{\left< m_{\beta\beta}\right>}{m_e}$, with $\left< m_{\beta\beta}\right>$ representing the 
effective Majorana neutrino mass, and $m_e$ the electron mass. $\eta^L_{N_R}$, $\eta^R_{N_R}$ are the heavy neutrino parameters with left-handed and right-handed currents, respectively~\cite{Horoi2013,Barry2013}, 
$\eta_{\tilde{q}}$, $\eta_{\lambda'}$ are $\cancel{\cal{R}}_p$ SUSY LNV parameters~\cite{Vergados2012}, 
$\eta_{\lambda}$, and $\eta_{\eta}$ are parameters for the so-called ''$\lambda-$'' and ''$\eta-$mechanism'', respectively~\cite{Barry2013}. 
$M_{0\nu}$, $M_{0N}$, are the light and the heavy neutrino exchange NME, $M_{\tilde{q}}$, $M_{\lambda'}$ are the $\cancel{\cal{R}}_p$ SUSY NME, and
$X_{\lambda}$ and $X_{\eta}$ denote combinations of NME and other PSF ($G_{02}-G_{09}$) corresponding to the the $\lambda-$mechanism involving right-handed leptonic and right-handed hadronic currents, and the $\eta-$mechanism with right-handed leptonic and left-handed hadronic currents, respectively~\cite{HoroiNeacsu2016prd}. The heavy neutrino exchange contribution to the amplitude in Eq. \ref{lrsm-hl} proportional with $M_N$ assumes that the heavy neutrino masses are larger than $\approx\ 1$ GeV and the information about their mass is included into the couplings\cite{HoroiNeacsu2016prd}.

In Table~\ref{tab-psf} we present the $Q^{0\nu}_{\beta\beta}$ values, the most recent experimental half-life limits from the indicated references, and the nine PSF for $0\nu\beta\beta$ transitions to ground states of the daughter nucleus for five isotopes currently under investigation. 
The PSF were calculated using a new effective method described in great detail in Ref.~\cite{HoroiNeacsu2016psf}. $G_{01}$ values were calculated with a screening 
factor ($s_f$) of 94.5, while for $G_{02}-G_{09}$ we used $s_f=92.0$ that was shown to provide results very close to those of Ref.~\cite{Stefanik2015}.
We note that the $^{82}$Se experimental half-life used here and throughout this analysis is preliminary~\cite{Waters-2016}. However, we believe that this limit is valid and that it may get improved.

In Ref.~\cite{HoroiNeacsu2016prd} we show how one could disentangle contributions form different mechanisms using two-electron angular and energy distributions, as well as half-life data from several isotopes. Here, we consider the case where one mechanism dominates, more explicitly, one single term in the decay amplitude of Eq. (\ref{lrsm-hl}).
Table~\ref{tab-etas} shows the shell model values of the NME that enter Eq. (\ref{lrsm-hl}). The light and heavy neutrino-exchange
NME, $M_{0\nu}$ and $M_{0N}$, are taken from Ref.~\cite{NeacsuHoroi2015} that describes their formalism and calculation. 
$M_{\tilde{q}}$ and $M_{\lambda^\prime}$ are calculated using the description in Eq. (150) and Eq. (155), respectively, of Ref.~\cite{Vergados2012}.
$X_{\lambda}$ and $X_{\eta}$ are adapted from $C_4$ and $C_5$ of Eq. (3.5.15d) and Eq. (3.5.15e), respectively, in Ref.~\cite{Doi1985} multiplied by $M_{GT}/G_{01}$ to fit the factorization of Eq. (\ref{lrsm-hl}).
All NME used in this paper 
 were calculated using the interacting shell model (ISM) approach~\cite{Blennow2010,Horoi2013,HoroiBrown2013,SenkovHoroiBrown2014,SenkovHoroi2013,BrownHoroiSenkov2014,NeacsuHoroi2015} (see Ref.~\cite{NeacsuHoroi2015} for a review), and include short-range-correlation effects based on the CD-Bonn parametrization~\cite{HoroiStoica2010}, finite-size effects~\cite{Vergados2012} and, when appropriate, optimal closure energies~\cite{Senkov2016} (see Appendix for more details). 

The upper limits for corresponding LNV parameters extracted from lower limits of the half-lives under the assumption that only one term in the amplitude dominates, are also presented in Table~\ref{tab-etas}.
There are a few other QRPA~\cite{Doi1985,Muto1989,Faessler1998-jpg,Vergados2012,Stefanik2015} and ISM~\cite{Retamosa1995,Caurier1996,Horoi2013,Neacsu2014constraints} results in the literature that were obtained within the framework of the LRSM and SUSY.
However, some of the extracted LNV parameters rely on some older half-life limits.

\begin{table}
 \caption{The NME that appear in Eq. (\ref{lrsm-hl})  for the five nuclei of current experimental interest, and the corresponding LNV parameters extracted under the assumption that only one dominates.}
 \begin{tabular}{@{}l@{}l@{}ccccc@{}}\label{tab-etas} 
&			&$^{48}$Ca\ \ \ &$^{76}$Ge\ \ \ &$^{82}$Se\ \ \ &$^{130}$Te\ \ \ &$^{136}$Xe \\ \hline
&$M_{0\nu}$		&$1.03$		&$3.64$		&$3.41$		&$1.93$	   	&$1.75$	\\
&$M_{0N}$ 		&$75.5$		&$202$		&$187$		&$136$		&$122$	\\
&$M_{\tilde{q}}	$	&$107$		&$339$		&$320$		&$185$		&$169$	\\
&$M_{\lambda^\prime}$	&$370$		&$619$		&$570$		&$415$		&$366$	\\
&$X_{\lambda}$		&$2.11$		&$4.12$		&$5.68$		&$2.81$		&$2.48$	\\
&$X_{\eta}$		&$246$		&$794$		&$725$		&$517$		&$467$	\\ \hline
$10^{6}\cdot$&$\left|\eta_{0\nu}\right|$&$27.1$	&$0.49$	&$3.64$		&$1.36$		&$0.28$	\\
$10^{9}\cdot$&$\left|\eta_{0N}\right|$	&$370.6$	&$8.83$	&$66.3$	&$19.2$		&$4.01$	\\
$10^{9}\cdot$&$\left| \eta_{\tilde{q}}\right|$&$260$	&$5.26$	&$38.8$	&$14.1$		&$2.91 $	\\
$10^{9}\cdot$&$\left| \eta_{\lambda^\prime}\right|$&$75.7$&$2.88$&$21.8$&$6.29$		&$1.34 $	\\
$10^{7}\cdot$&$\left| \eta_\lambda \right|$&$133$&$4.32$&$21.87$	&$9.29$		&$1.98$ \\
$10^{9}\cdot$&$\left| \eta_\eta \right|$	&$114$	&$2.25$	&$17.1$	&$5.05$		&$1.05 $	\\
\end{tabular}
 \end{table} 

 \begin{figure}
{\centering
\hfill { \put(0,40){\makebox(0,0){ }}} \hfill
    \subfloat[The generic $0\nu\beta\beta$ decay diagram at the quark-level.]
   {\includegraphics[width=0.22\textwidth]{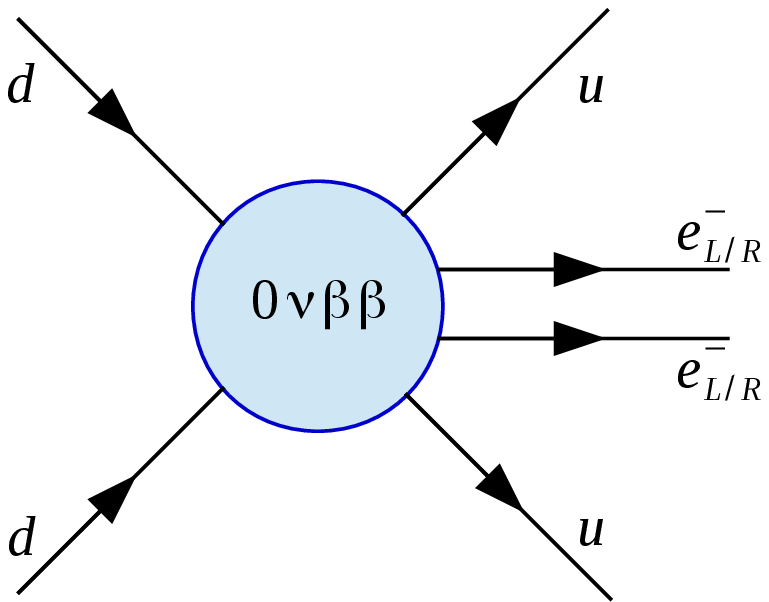} \label{subfig-generic}  } 
\hfill { \put(0,40){\makebox(0,0){=}}} \hfill
    \subfloat[Light left-handed neutrino exchange diagram.]
    {\includegraphics[width=0.22\textwidth]{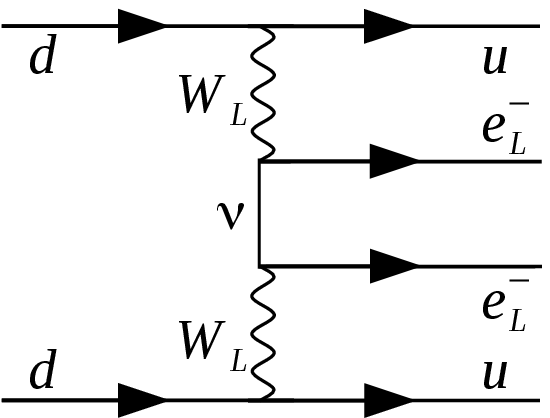} \label{subfig-light}      } \\
\hfill { \put(0,40){\makebox(0,0){+}}} \hfill
     \subfloat[The long-range part of the $0\nu\beta\beta$ diagram.]
    {\includegraphics[width=0.22\textwidth]{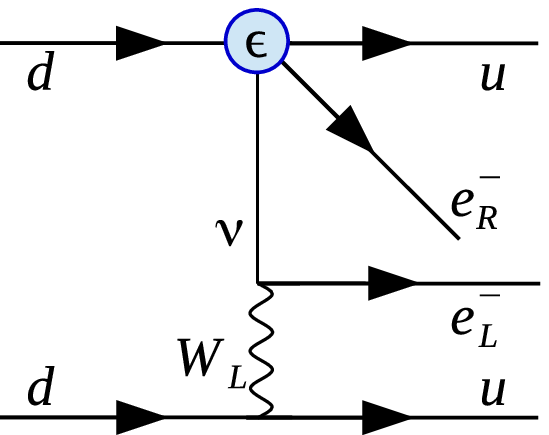}\label{subfig-longrange} } 
\hfill { \put(0,40){\makebox(0,0){+}}} \hfill
    \subfloat[The short-range part of the $0\nu\beta\beta$ diagram.]
    {\includegraphics[width=0.22\textwidth]{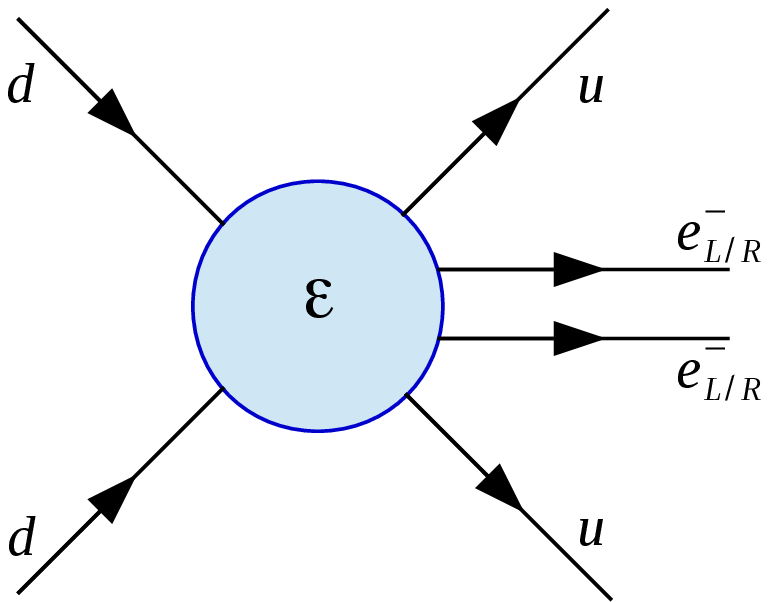}\label{subfig-shortrange}}       
    \caption{The $0\nu\beta\beta$ decay process diagrams: (\ref{subfig-generic}) presents the generic description of the process,
(\ref{subfig-light}) shows the most studied case in the literature, that of the light left-handed neutrino exchange, (\ref{subfig-longrange}) 
is the long-range component of the $0\nu\beta\beta$ decay diagram, while (\ref{subfig-shortrange}) displays the short-range part.}
    \label{fig-diagrams} }
\end{figure}
\section{Effective field theory approach to neutrinoless double-beta decay} \label{section-eft}

A more general approach is based on the effective field theory extension of the Standard Model. 
The analysis based on the BSM contributions to the effective field theory is more desirable, because it does not rely on specific models, 
and their parameters could be extracted/constrained by the existing $0\nu\beta\beta$ data, and by data from LHC and other experiments. 
In fact, the models considered in section~\ref{section-mechanisms} always lead to a subset of terms in the low-energy ($\sim$ 200 MeV) effective field theory Lagrangian. 
Here we consider all the terms in the Lagrangian allowed by the symmetries. Some of the couplings will correspond to the model couplings in Eq. (\ref{lrsm-hl}), 
but they might have a wider meaning. Others are new, not corresponding to specific models.

At the quark-level, we present in Figure~\ref{fig-diagrams} the generic $0\nu\beta\beta$ Feynman diagrams contributing to the $0\nu\beta\beta$ process. We consider contributions 
coming from the light left-handed Majorana neutrino (Fig.~\ref{subfig-light}), a long-range part coming from the low-energy four-fermion charged-current interaction (Fig.~\ref{subfig-longrange}),
and a short-range part (Fig.~\ref{subfig-shortrange}).

We treat the long-range component of the $0\nu\beta\beta$ diagram as two point-like vertices at the Fermi scale, which exchange a light neutrino.
In this case, the dimension 6 Lagrangian can be expressed in terms of effective couplings~\cite{Deppisch2012}:
\begin{equation} \label{lag-longrange}
 \mathcal{L}_6=\frac{G_F}{\sqrt{2}}\left[ j^\mu_{V-A} J^\dagger_{V-A,\mu} + \sum^{*}_{\alpha,\beta} \epsilon_\alpha^\beta j_\beta J^\dagger_\alpha \right] ,
\end{equation}
where $J^\dagger_\alpha=\bar{u}\mathcal{O}_\alpha d$ and $j_\beta=\bar{e}\mathcal{O}_\beta \nu$ are hadronic and leptonic Lorentz currents, respectively.
The definitions of the $\mathcal{O}_{\alpha,\beta}$ operators are given in Eq. (3) of Ref.~\cite{Deppisch2012}. The LNV parameters are $\epsilon_\alpha^\beta= \{ \epsilon^{V+A}_{V-A}, \ \epsilon^{V+A}_{V+A}, \ \epsilon^{S+P}_{S\pm P}, \ \epsilon^{TR}_{TL},\ \epsilon^{TR}_{TR} \} $.
The ''*'' symbol indicates that the term with $\alpha=\beta=(V-A)$ is explicitly taken out of the sum. However, the first term in Eq. (\ref{lag-longrange}) still entails BSM physics through the dimension-5 operator responsible for the Majorana neutrino mass (see also section~\ref{discussions}).
Here $G_F=1.1663787\times 10^{-5}$ GeV$^{-2}$ denotes the Fermi coupling constant. 

As already mentioned, some of these couplings play the same role as some of the model couplings listed in Eq. (\ref{lrsm-hl}), but they have more general meaning here. For example, $\epsilon^{V+A}_{V-A}$ play the same role as $\eta_{\eta}$ and $\epsilon^{V+A}_{V+A}$ play the same role as $\eta_{\lambda}$ in the effective Lagrangian associated to models.

In the short-range part of the diagram presented in Fig.~\ref{subfig-shortrange} we consider the interaction to be point-like. 
Expressing the general Lorentz-invariant Lagrangian in terms of effective couplings~\cite{Pas2001}, we get:
\begin{flalign}
\nonumber \mathcal{L}_9=&\frac{G_F^2}{2 m_p} \Bigl[ \varepsilon_1 JJj + \varepsilon_2 J^{\mu\nu} J_{\mu\nu}j + \varepsilon_3 J^{\mu} J_{\mu}j \Bigr. \\
  +& \Bigl. \varepsilon_4 J^{\mu} J_{\mu\nu}j^\nu + \varepsilon_5 J^{\mu} Jj_\mu \Bigr] , 
\label{lag9}
\end{flalign}
with the hadronic currents of defined chirality $J=\bar{u}(1\pm\gamma_5)d$, $J^\mu=\bar{u}\gamma^\mu(1\pm\gamma_5)d$, $J^{\mu\nu}=\bar{u}\frac{i}{2}[\gamma^\mu,\gamma^\nu](1\pm\gamma_5)d$, leptonic currents $j=\bar{e}(1\pm\gamma_5)e^C$, $j^\mu=\bar{e}\gamma^\mu(1\pm\gamma_5)e^C$, and $\varepsilon_\alpha^\beta=\varepsilon_\alpha^{xyz}=\{ \varepsilon_{1}, \ \varepsilon_{2}, \ \varepsilon_{3}^{LLz(RRz)}, \ \varepsilon_{3}^{LRz(RLz)}, \ \varepsilon_{4}, \ \varepsilon_{6} \}$.
These parameters have dependence on the chirality of the hadronic and the leptonic currents involved, with $xyz=L/R, L/R, L/R$.
In the case of $\varepsilon_3$, one can distinguish between different chiralities, thus we express them separately as
$\varepsilon_{3}^{LLz(RRz)}$ and $ \varepsilon_{3}^{LRz(RLz)}$.
\begin{figure*}    
    \centering
      \begin{minipage}{\linewidth}
    \subfloat[The generic $0\nu\beta\beta$ decay process nucleon-level diagram.]
   {\includegraphics[width=0.22\textwidth]{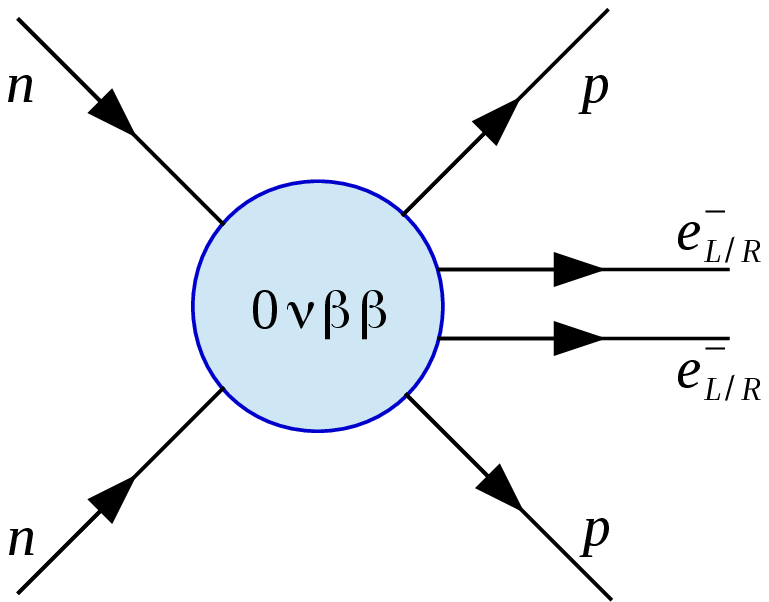} \label{subfig-ngeneric}  } 
     \hfill { \put(0,40){\makebox(0,0){=}}} \hfill
    \subfloat[Light left-handed neutrino exchange diagram.]
    {\includegraphics[width=0.22\textwidth]{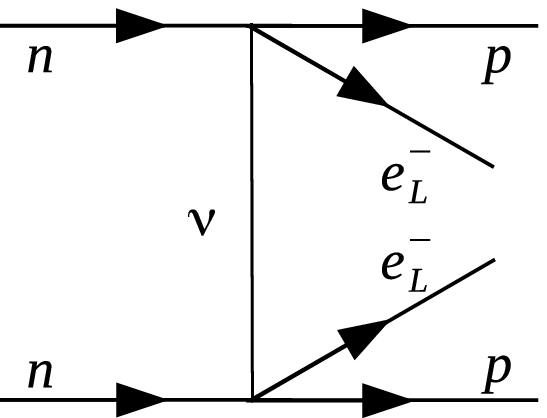} \label{subfig-nlight}      } 
     \hfill { \put(0,40){\makebox(0,0){+}}} \hfill
     \subfloat[The The nucleon-nucleon long-range $\mathcal{L}_6$ mode.]
    {\includegraphics[width=0.22\textwidth]{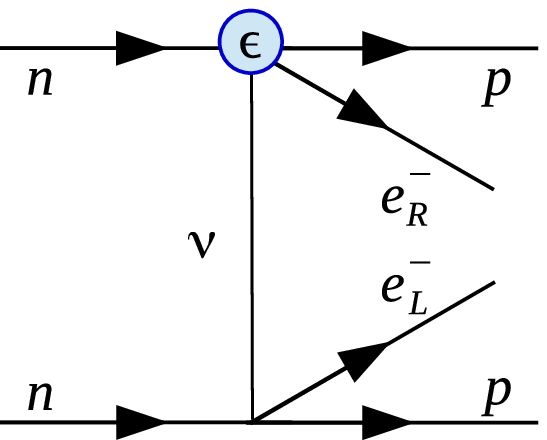}\label{subfig-nlongrange} } 
     \hfill { \put(0,40){\makebox(0,0){+}}} \hfill
    \subfloat[The nucleon-nucleon short-range part of the $0\nu\beta\beta$ diagram.]
    {\includegraphics[width=0.22\textwidth]{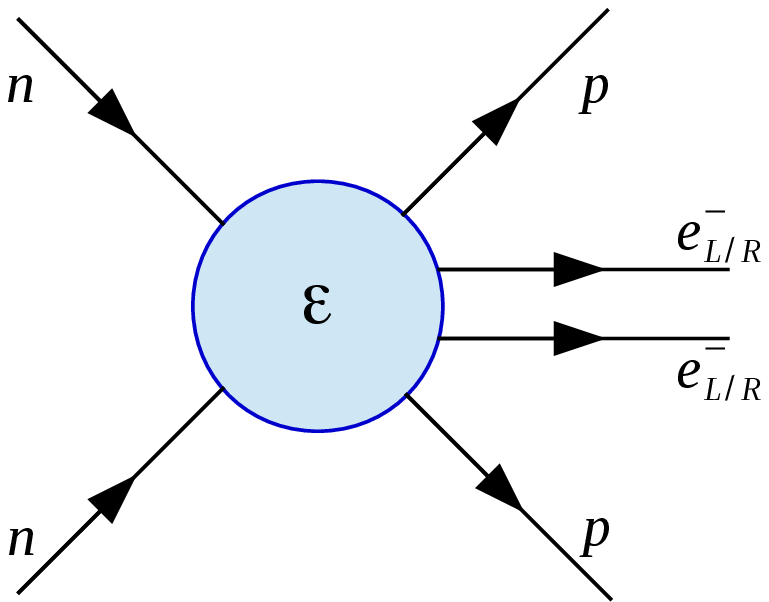} \label{subfig-nheavy}      } 
      \end{minipage} 
     
      \begin{minipage}{\linewidth}
    \hspace*{.22\textwidth} 
     \hfill { \put(0,40){\makebox(0,0){+}}} \hfill
     \subfloat[The pion-neutrino long-range diagram.]
    {\includegraphics[width=0.22\textwidth]{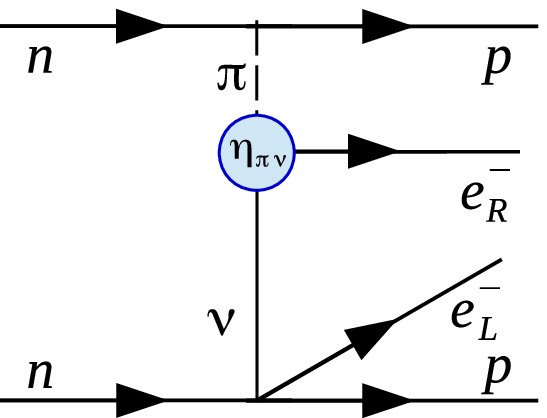}\label{subfig-npion_nu} } 
     \hfill { \put(0,40){\makebox(0,0){+}}} \hfill
     \subfloat[The one-pion long-range diagram.] 
    {\includegraphics[width=0.22\textwidth]{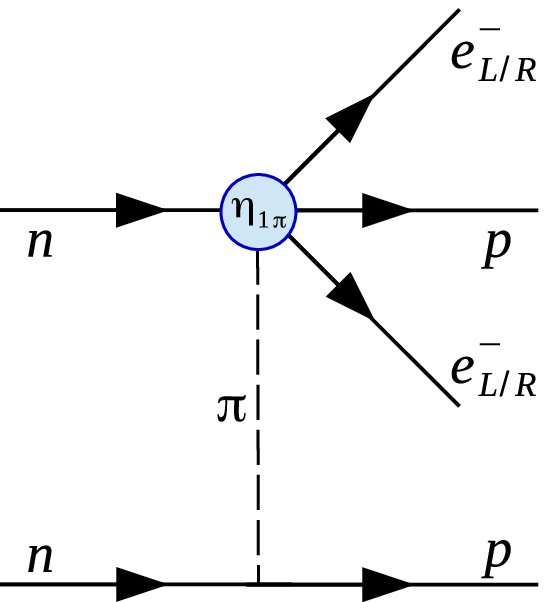}\label{subfig-nonepion} } 
     \hfill { \put(0,40){\makebox(0,0){+}}} \hfill
     \subfloat[The two-pion long-range diagram.]
    {\includegraphics[width=0.22\textwidth]{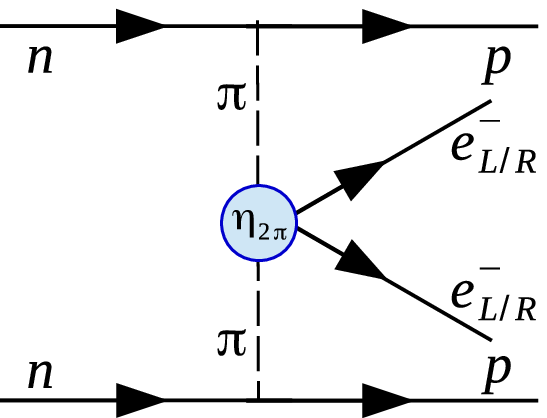}\label{subfig-ntwopion} } 
      \end{minipage}
\caption{Similar to Fig.\ref{fig-diagrams}, we present the nucleon-level diagrams of $0\nu\beta\beta$ decay process : 
(\ref{subfig-ngeneric}) presents the generic description of the process,
(\ref{subfig-nlight}) shows the light left-handed neutrino exchange, (\ref{subfig-nlongrange}) is the long-range component, 
Subfigure~\ref{subfig-nheavy} shows the short-range contribution.
On the second line, (\ref{subfig-npion_nu}) is the pion-neutrino component, 
(\ref{subfig-nonepion}) is the one-pion long-range contribution of the $\cancel{\cal{R}}_p$ SUSY induced $0\nu\beta\beta$ diagram,
 and (\ref{subfig-ntwopion} presents the two-pion long-range contribution of the $\cancel{\cal{R}}_p$ SUSY induced $0\nu\beta\beta$. The effective
 couplings $\eta_{1\pi}$ and $\eta_{2\pi}$ are related to Eq. (\ref{cal_nmes2prime}) as $\eta_{1\pi}=c_{1\pi}\eta_{\pi N}$ and $\eta_{2\pi}=c_{2\pi}\eta_{\pi N}$.}
\label{fig-ndiagrams} 
 \end{figure*}

The contribution of the diagrams~\ref{subfig-light} and~\ref{subfig-longrange} to the $0\nu\beta\beta$ decay amplitude is proportional to the time-ordered product of two effective $\mathcal{L}_6$ Lagrangians~\cite{Deppisch2012},
\begin{flalign}
\nonumber \! \! T(\mathcal{L}_6^{(1)}\mathcal{L}_6^{(2)})&=\frac{G^2_F}{2}T\left[ j_{V-A}J^\dagger_{V-A}j_{V-A}J^\dagger_{V-A} \right.\\
 &+ \left. \epsilon_\alpha^\beta j_\beta J^\dagger_\alpha j_{V-A} J^\dagger_{V-A}
+\epsilon_\alpha^\beta \epsilon_\gamma^\delta j_\beta J^\dagger_\alpha j_\delta J^\dagger_\gamma \right], 
 \end{flalign}
while the contribution of the diagram~\ref{subfig-shortrange} is proportional to $\mathcal{L}_9$.

However, when calculating the $0\nu\beta\beta$ half-life it is necessary to identify the contributions  corresponding to different hadronization prescriptions.
Figure~\ref{fig-ndiagrams} shows the nucleon-level diagrams in a similar way to Figure~\ref{fig-diagrams}.
The first 3 contributions, Figs.~\ref{subfig-nlight},~\ref{subfig-nlongrange}, and~\ref{subfig-nheavy} are similar to the corresponding amplitudes at the quark level (see Fig.~\ref{fig-diagrams}). In addition to these contributions that were also considered in Ref.~\cite{Deppisch2012}, here we also include the long range diagrams that involve pion(s) exchange, Figs.~\ref{subfig-npion_nu},~\ref{subfig-nonepion}, and~\ref{subfig-ntwopion}. These diagrams were considered before as contributing to the $0\nu\beta\beta$ decay rate, but in the context of $\cancel{\cal{R}}_p$ SUSY mechanism. For example, the diagram~\ref{subfig-npion_nu} was considered to describe the contribution of the squark-exchange mechanism~\cite{Kovalenko2008}, and the diagrams 
\ref{subfig-nonepion} and~\ref{subfig-ntwopion} were considered to describe the contribution of the gluino exchange mechanism~\cite{Wodecki1999}. One should also mention that the diagram~\ref{subfig-ntwopion} was also considered in Refs.~\cite{Prezeau2003,Peng2016}, but its contribution to the $0\nu\beta\beta$ half-life was estimated differently, and cannot be directly compared to the other contributions analyzed here. 

After hadronization (see Fig.~\ref{fig-ndiagrams}), the extra terms in the Lagrangian require the knowledge of  23 individual NME~\cite{Hirsch1996,Deppisch2012,Pas1999,Pas2001,Deppisch2015-PRD92,Vergados2012}. 
We can write the half-life in a factorized compact form 
\begin{flalign} \label{flifetime}
 \left[ T^{0\nu}_{1/2}\right]^{-1}\! \! \! \! =&g_{A}^4 \left[ \sum_{i} \left| \mathcal{E}_{i} \right|^2 {\cal{M}}_{i}^2 + \textnormal{Re} \left[ \sum_{i \ne j} \mathcal{E}_{i}\mathcal{E}_{j} {\cal{M}}_{ij}\right] \right].
\end{flalign}
Here, the $\mathcal{E}_{i}$ contain the neutrino physics parameters, with $\mathcal{E}_{1}=\eta_{0\nu}$ representing the exchange of light left-handed neutrinos 
corresponding to Fig.~\ref{subfig-nlight}, $\mathcal{E}_{2-7}= \{ \epsilon^{V+A}_{V-A}, \ \epsilon^{V+A}_{V+A}, \ \epsilon^{S+P}_{S\pm P}, \ \epsilon^{TR}_{TL},\ \epsilon^{TR}_{TR},\ \eta_{\pi\nu}\} $ are the long-range LNV parameters appearing in 
Figs.~\ref{subfig-nlongrange} and~\ref{subfig-npion_nu}, and $\mathcal{E}_{8-15}=\{ \varepsilon_{1}, \ \varepsilon_{2}, \ \varepsilon_{3}^{LLz(RRz)}, \ \varepsilon_{3}^{LRz(RLz)}, \ \varepsilon_{4}, \ \varepsilon_{5},\ \eta_{1\pi},\ \eta_{2\pi} \}$ 
denote the short-range LNV parameters at the quark level involved in the diagrams of Fig.~\ref{subfig-nheavy},~\ref{subfig-nonepion},~\ref{subfig-ntwopion}. The rational for including the $\eta_{\pi\nu}$ in the same class with the LNV entering the quark-level long range diagrams is that Ref.~\cite{Kovalenko2008} indicates that $\eta_{\pi\nu}$ is proportional to $\epsilon^{TR}_{TR}$ (see Section~\ref{subsection-long} below). In the same vein, Ref.~\cite{Wodecki1999} indicates that  $\varepsilon_{1}$ and  $\varepsilon_{2}$ are proportional to a combination of $\eta_{1\pi}$ and $\eta_{2\pi}$ (see Section~\ref{subsection-short} below). Therefore the $\eta_{1\pi}$ and $\eta_{2\pi}$ were included in the list LNV couplings associated with quark-level short-range diagrams. Contributions of pion-exchange diagrams similar to those of Figs.~\ref{subfig-nonepion} and~\ref{subfig-ntwopion} are also included in the so called  "higher order term in nucleon currents" \cite{Vergados2012}. However, they are constrained by PCAC, and are only included in light-neutrino exchange contribution of diagram~\ref{subfig-ngeneric}. This contribution changes the associated NME by only 20\%. Therefore, we conclude that this does not represent a serious double counting issue.

Following Refs.~\cite{Pas1999,Pas2001,Deppisch2012,Vergados2012}, we write ${\cal{M}}_{i}^2$ as combinations of NME described in Eqs. (\ref{cal_nmes},~\ref{cal_nmes1},~\ref{cal_nmes1prime},~\ref{cal_nmes2}, and~\ref{cal_nmes2prime}) (see also Eq.(\ref{eq-Ma}) in the Appendix for the individual NME) and integrated 
PSF~\cite{HoroiNeacsu2016psf} denoted with $G_{01}-G_{09}$. Our values of the PSF are presented in Table~\ref{tab-psf}. 
In some cases the interference terms $\mathcal{E}_{i}\mathcal{E}_{j} {\cal{M}}_{ij}$ are small~\cite{Ahmed2017} and 
can be neglected, but not all of them. In Ref.~\cite{HoroiNeacsu2016prd} we analyzed a subset of terms contributing to the half-life formula, Eq. (\ref{lrsm-hl}) originating from the left-right symmetric model. In that restrictive case we showed that one can disentangle different contributions to the $0\nu\beta\beta$ decay process using two-electron angular and energy distributions as well as half-lives of two selected isotopes. Obviously, this number of observables is not enough to extract all coupling appearing in the effective field theory Lagrangian. However, they can be used to constrain these couplings, thus adding to the information extracted from the Large Hadron Collider and other related experiments. 
Here we attempt to extract these couplings assuming that only one of them can have a dominant contribution to the half-life, Eq. (\ref{flifetime}). We call this approach ``on-axis``.
Considering the ``on-axis`` approach to extracting limits of the LNV parameters, the interference terms are neglected in our analysis. 
In the following, we extract the ``on-axis`` upper limits of these parameters using the most recent experimental half-lives lower limits, as presented in Table~\ref{tab-psf}.

\section{Experimental limits on the BSM LNV couplings} \label{results}
To obtain experimentally constrained upper limits of the effective LNV couplings one needs experimental half-life lower limits, accurate calculations of the PSF,
together with reliable NME results calculated using 
nuclear structure methods tested to correctly describe the experimental nuclear structure data available for the nuclei involved.
 We split our analysis of the LNV parameters into three subsections corresponding the exchange of light left-handed Majorana neutrinos, the LNV couplings entering the remaining quark-level long-range diagrams, and the LNV couplings entering the quark-level short-range diagrams.

\subsection{The exchange of light left-handed neutrinos} \label{subsection-mass}
Most studies in the literature have considered just the case where only the exchange of light left-handed Majorana neutrinos contribute to the $0\nu\beta\beta$ decay process, presented in 
Figs.~\ref{subfig-light} and~\ref{subfig-nlight}. Therefore, one can easily find calculations of NME and PSF for this scenario. 
Considering this case, we reduce the half-life equation to:
\begin{equation}
 \left[ T^{0\nu}_{1/2}\right]^{-1}=g_{A}^4 \left| \eta_{0\nu}\right|^2 {\cal{M}}_{0\nu}^2 ,
\label{thmass}
\end{equation}
where $g_A=1.27$, ${\cal{M}}^2_{0\nu}$ contains the coefficients containing combinations of NME and PSF (see Eq. (\ref{cal_nmes}) below). $\eta_{0\nu}=\frac{\left< m_{\beta\beta} \right>}{m_e}$,  where $m_e$ is the electron mass and ${\left< m_{\beta\beta} \right>}$ represents the effective Majorana neutrino mass described as~\cite{Rodejohann2012}:
\begin{equation}
 \left< m_{\beta\beta}\right>= \left| \sum_{j=1}^3 U^2_{ej}m_j \right|.
\end{equation}
Here $U_{ej}$ are the PMNS mixing matrix elements~\cite{pmns,upmns} and the summation is performed over all the three light neutrino mass eigenstates $m_j$. Also in Eq. (\ref{thmass})
\begin{equation} \label{cal_nmes}
{\cal{M}}^2_{0\nu}= G_{01} \left[ M_{GT}-\left(\frac{g_V}{g_A}\right)^2 M_F + M_T \right]^2,
\end{equation}
\noindent where $g_V=1$ is the vector coupling constant, $g_A=1.27$ is the axial coupling constant, and $G_{01}$ is the phase-space factor. The three NME, $M_{GT}$, $M_F$, and $M_T$ (shown in Table~\ref{tab-nme-light}) correspond to the Gamow-Teller, Fermi and Tensor transition operators, respectively, and are described in the Appendix. 
All the NME listed in the tables of the Appendix have the correct signs relative to that of $M_{GT}$, which is chosen to be positive. The ${\cal{M}}^2$ coefficients correctly include these relative signs, but the overall sign of the $\cal{M}$ in Eqs. (\ref{cal_nmes},~\ref{cal_nmes1},~\ref{cal_nmes1prime},~\ref{cal_nmes2}, and~\ref{cal_nmes2prime}) is lost due to squaring.
 \begin{table}
 \caption{The first line shows values of the ${\cal{M}}_{0\nu}^2$ coefficients containing combinations of NME and PSF, and the second line presents the extracted neutrino physics parameter $\left|\eta_{0\nu}\right|$ for the most studied case, 
 assuming only the exchange of light left-handed Majorana neutrinos. }
 \begin{tabular}{l@{}lccccc}\label{tab-mrond-light} 
			  &		  &$^{48}$Ca\ \ \ &$^{76}$Ge\ \ \ &$^{82}$Se\ \ \ &$^{130}$Te\ \ \ &$^{136}$Xe \\ \hline
${\cal{M}}_{0\nu}^2$	  &$\cdot 10^{14}$&$2.61$	&$3.01$		&$11.6$		  &$5.23$	   &$4.41$	\\
$\left|\eta_{0\nu}\right|$&$\cdot 10^{6} $&$27.1$	&$0.49$		&$3.6$		  &$1.36$	   &$0.28$	
\end{tabular}
 \end{table} 

In Table~\ref{tab-mrond-light} we present the ${\cal{M}}_{0\nu}^2$ values and their corresponding $\eta_{0\nu}$ limits.
We find the lowest upper-limit of this parameter for $^{136}$Xe, which leads to a limit for the Majorana neutrino mass $\left< m_{\beta\beta}\right>\sim 140$ meV.

\subsection{The long-range effective LNV couplings} \label{subsection-long}

Investigating the ``on-axis`` LNV parameters of the diagram of Fig.~\ref{subfig-nlongrange}, the half-life is factorized as:
\begin{equation} \label{flongrange}
 \left[ T^{0\nu}_{1/2}\right]^{-1}=g_{A}^4 \left[ \left| \epsilon_\alpha^\beta \right|^2 {\cal{M}}_{\alpha\beta}^2 \right],
 \end{equation}
with $\epsilon_\alpha^\beta= \{ \epsilon^{V+A}_{V-A}, \ \epsilon^{V+A}_{V+A}, \ \epsilon^{S+P}_{S\pm P}, \ \epsilon^{TR}_{TL},\ \epsilon^{TR}_{TR} \} $. Here and below the $\alpha\beta$ combination corresponds to some index $i$ in Eq. (\ref{flifetime}), as described in the definition of $\mathcal{E}_i$ after Eq. (\ref{flifetime}).
Following the formalism presented in Refs.~\cite{Deppisch2012,Pas1999,Doi1985} and including the $G_{01}-G_{09}$ PSF, we write the long-range coefficients containing combinations of NME and PSF as:
\begin{subequations}
\label{cal_nmes1}
\begin{flalign}
\nonumber {\cal{M}}_{V+A/V-A}^2	& = G_{02}{\cal{M}}^2_{2+}\! \! -\frac{2}{9}G_{03}{\cal{M}}_{1-}{\cal{M}}_{2+}\! \! +\frac{1}{9}G_{04}{\cal{M}}^2_{1-}\\
				& -G_{07}M_P M_R+G_{08}M^2_P+G_{09}M^2_R,\\
\nonumber {\cal{M}}_{V+A/V+A} ^2& = G_{02}{\cal{M}}^2_{2-}\\
				& -\frac{2}{9}G_{03}{\cal{M}}_{1+}{\cal{M}}_{2-}  +\frac{1}{9}G_{04}{\cal{M}}^2_{1+}, \\
\nonumber {\textmd {with\ }}{\cal{M}}_{1\pm} & = M_{GTq}\pm 3 \left(\frac{g_V}{g_A}\right)^2M_{Fq} - 6 M_{Tq} \\ 
\nonumber {\textmd{and}} \ {\cal{M}}_{2\pm} & =M_{GT\omega} \pm \left(\frac{g_V}{g_A}\right)^2 M_{F\omega} -\frac{1}{9}{\cal{M}}_{1\pm}, \\
{\cal{M}}_{S+P/S\pm P}^2	& = G_{01} \left[\frac{F^{(3)}_P}{R m_e g_A} \left( \frac{1}{3} M_{GT'} + M_{T'} \right) \right]^2 \! \! \! \! , \\
\nonumber{\cal{M}}_{TR/TR} ^2	& = G_{01} \left[ \frac{4T^{(3)}_1 g_V(1-(\mu_p-\mu_n)) }{R m_e g_A^2} \right. \\
				&\times  \left. \left( M_{T'}- \frac{2}{3} M_{GT'} \right) \right]^2, \\
\nonumber{\cal{M}}_{TR/TL} ^2		& = G_{01} \left[ \frac{g_V \left( 8\hat{T}^{(3)}_2-4T^{(3)}_1\right) } {R m_e g_A^2} M_{F'} \right.  \\
				& - \left. \frac{4T_1^{(3)}2m_p}{g_A R^2 m_e m^2_\pi}\left(M_{T''}+\frac{1}{3}M_{GT''} \right) \right]^2  \! \! \! \! .
\end{flalign}
\end{subequations}
In these equations, $R=1.2A^{1/3}$ fm is the nuclear radius, $m_e=0.511$ MeV is the electron mass, $m_\pi=139$ MeV is the pion mass, $m_p=938$ MeV is the proton mass, $(\mu_p-\mu_n)\simeq 3.7$,
and the parameters $F_P^{(3)}=4.41$, $T_1^{(3)}=1.38$, $\hat{T}_2^{(3)}=-4.54$ are taken from Ref.~\cite{Adler1975} where they have been calculated using the MIT bag model.
Detailed expressions for the individual $M_\alpha$ (with $\alpha=$ $GTq$, $Fq$, $Tq$, $GT\omega$, $F\omega$, $P$, $R$, $GT'$, $F'$, $T'$, $GT''$, $T''$) are found in the Appendix. 

It is possible to obtain another limit for $\epsilon_{TR}^{TR}$ by considering a different hadronization procedure~\cite{Kovalenko2008} depicted in Fig.~\ref{subfig-npion_nu}, where our $\eta_{\pi\nu}$ plays the same role as $\eta^{11}_{(q)LR}$ in Eq.(22) of Ref.~\cite{Kovalenko2008}. In this case we can obtain an alternative value for $\epsilon_{TR}^{TR}$, $\tilde{\epsilon}_{TR}^{TR}=\eta_{\pi\nu} /8$.
\begin{equation}
 \left[ T^{0\nu}_{1/2}\right]^{-1}=g_{A}^4 \left[ \left| 8 \ \tilde{\epsilon}_{TR}^{TR} \right|^2 {\cal{M}}_{\pi\nu}^2 \right],
\end{equation}
with 
\begin{equation}\label{cal_nmes1prime}
 {\cal{M}}_{\pi\nu}^2 = G_{01} \left[ M_{GT{\pi\nu}} + M_{T{\pi\nu}} \right]^2.
\end{equation}
The $M_{GT{\pi\nu}}$ and $M_{T{\pi\nu}}$  are the same NME as $M_{GT(\tilde{q})}$ and $M_{GT(\tilde{q})}$
in Eq.(155) of Ref.\cite{Vergados2012} (also described in the Appendix). 
  \begin{table}[hb]
\caption{The ${\cal{M}}_{\alpha\beta}^2$ values for the long-range part of the $0\nu\beta\beta$ decay process.}
 \begin{tabular}{r@{}lccccc}\label{tab-mrond-lr}
		&			&$^{48}$Ca\ \ \	&$^{76}$Ge\ \ \	&$^{82}$Se\ \ \	&$^{130}$Te\ \ \ &$^{136}$Xe \\ \hline
$10^{9}\cdot$	&${\cal{M}}_{V+A/V-A}^2$&$1.49$		&$1.44$		&$5.24$		&$3.76$		&$3.17$	\\
$10^{13}\cdot$	&${\cal{M}}_{V+A/V+A}^2$&$1.09$		&$0.39$		&$3.21$		&$1.11$		&$0.89$	 \\
$10^{10}\cdot$&${\cal{M}}_{S+P/S\pm P}^2$&$1.62$	&$1.17$		&$4.20$		&$4.73$		&$4.18$ \\
$10^{8}\cdot$	&${\cal{M}}_{TR/TR}^2$	&$6.07$		&$1.15$		&$4.43$		&$2.45$		&$1.95$ \\
$10^{7}\cdot$	&${\cal{M}}_{TR/TL}^2$	&$3.42$		&$0.25$		&$0.80$		&$1.74$		&$1.59$ \\ 
$10^{10}\cdot$	&${\cal{M}}_{\pi\nu}^2$	&$2.84$		&$2.62$		&$10.2$		&$4.85$		&$4.13$
\end{tabular}
 \end{table}

Table~\ref{tab-mrond-lr} shows our shell model ${\cal{M}}^2_{\alpha\beta}$ coefficients. We present our values for the long-range LNV parameters in Table~\ref{tab-epslr}, where $\tilde{\epsilon}^{TR}_{TR}$ represents the alternative limit for
$\epsilon^{TR}_{TR}$ that is obtained using ${\cal{M}}_{\pi\nu}$. With the exception of $^{48}$Ca, the  $\tilde{\epsilon}^{TR}_{TR}$ upper-limits are slightly
lower than those of $\epsilon^{TR}_{TR}$.
 \begin{table}
\caption{The ``on-axis`` values of the long-range LNV parameters $\epsilon^{\beta}_{\alpha}$. The last two lines present $\eta_{\pi\nu}$ and its corresponding $\tilde{\epsilon}^{TR}_{TR}$ limit. }
 \begin{tabular}{llllll}\label{tab-epslr}
					&$^{48}$Ca		&$^{76}$Ge		&$^{82}$Se		&$^{130}$Te		&$^{136}$Xe \\ \hline
$\left| \epsilon^{V+A}_{V-A} \right|$	&$1.1\cdot 10^{-7}$	&$2.2\cdot 10^{-9}$	&$1.7\cdot 10^{-8}$	&$5.1\cdot 10^{-9}$	&$1.1\cdot 10^{-9}$	\\
$\left| \epsilon^{V+A}_{V+A} \right|$	&$1.3\cdot 10^{-5}$	&$4.3\cdot 10^{-7}$	&$2.2\cdot 10^{-6}$	&$9.3\cdot 10^{-7}$	&$2.0\cdot 10^{-7}$ \\
$\left| \epsilon^{S+P}_{S\pm P} \right|$&$3.4\cdot 10^{-7}$	&$7.9\cdot 10^{-9}$	&$6.1\cdot 10^{-8}$	&$1.4\cdot 10^{-8}$	&$2.9\cdot 10^{-9}$ \\
$\left| \epsilon^{TR}_{TR} \right|$	&$1.8\cdot 10^{-8}$	&$7.9\cdot 10^{-10}$	&$5.9\cdot 10^{-9}$	&$2.0\cdot 10^{-9}$	&$4.2\cdot 10^{-10}$ \\ 
$\left| \epsilon^{TR}_{TL} \right|$	&$7.5\cdot 10^{-9}$	&$5.4\cdot 10^{-10}$	&$4.4\cdot 10^{-9}$	&$7.4\cdot 10^{-10}$	&$1.5\cdot 10^{-10}$ \\
$\left| \eta_{\pi\nu} \right|$		&$2.6\cdot 10^{-7}$	&$5.3\cdot 10^{-9}$	&$3.9\cdot 10^{-8}$	&$1.4\cdot 10^{-8}$	&$2.9\cdot 10^{-9}$  \\
$\left| \tilde{\epsilon}^{TR}_{TR} \right|$&$3.3\cdot 10^{-8}$	&$6.6\cdot 10^{-10}$	&$4.8\cdot 10^{-9}$	&$1.8\cdot 10^{-9}$	&$3.6\cdot 10^{-10}$ 
\end{tabular}
 \end{table} 

The shell model values for $M_\alpha$, with $\alpha=$ $GTq$, $Fq$, $Tq$, $GT\omega$, $F\omega$, $P$, $R$, $GT'$, $F'$, $T'$, $GT''$, $T''$, $GT\pi\nu$, $T\pi\nu$, are shown in Table~\ref{tab-nme-muto} of the Appendix.
 
\subsection{The short-range LNV couplings} \label{subsection-short}

Similar to the case of the long-range component, we extract the ``on-axis`` values of the short-range LNV parameters using the following expression for the half-life corresponding
to the diagram of Fig.~\ref{subfig-shortrange}:
\begin{equation} \label{fshortrange}
 \left[ T^{0\nu}_{1/2}\right]^{-1}=g_{A}^4 \left[ \left| \varepsilon_\alpha^\beta \right|^2 {\cal{M}}_{\alpha\beta}^2 \right],
\end{equation}
with $\varepsilon_\alpha^\beta=\{ \varepsilon_{1}, \ \varepsilon_{2}, \ \varepsilon_{3}^{LLz(RRz)}, \ \varepsilon_{3}^{LRz(RLz)}, \ \varepsilon_{4}, \ \varepsilon_{6} \}$.
The index $\beta=xyz$, with $xyz=L/R, L/R, L/R$, indicates the chirality of the hadronic and the leptonic currents. It is only possible to distinguish between the different chiralities in the case of $\varepsilon_3$ where we denote them explicitly as $\varepsilon_{3}^{LLz(RRz)}$ and $ \varepsilon_{3}^{LRz(RLz)}$.
For the other cases we omit this labeling.

Adapting the formalism of Ref.~\cite{Deppisch2012,Pas2001,Vergados2012}, we can write the coefficients containing combinations of NME and PSF as:
\begin{subequations}\label{cal_nmes2}
\begin{flalign} 
{\cal{M}}_{1} ^2	& = G_{01}\left[ \left( \frac{F_S^{(3)}}{g_A}\right)^2 M_{FN}\right]^2, \\
{\cal{M}}_{2} ^2	& = G_{01}\left[ \left( 8\frac{T_1^{(3)}}{g_A} \right)^2 M_{GTN} \right]^2, \\
\nonumber {\cal{M}}_{3LLz} ^2	& = {\cal{M}}_{3RRz} ^2 \\
			& = G_{01}\left[M_{GTN} - \left( \frac{g_V}{g_A}\right)^2 M_{FN} \right]^2,\\
\nonumber {\cal{M}}_{3LRz} ^2	& = {\cal{M}}_{3RLz} ^2 \\
			& = G_{01}\left[M_{GTN} + \left( \frac{g_V}{g_A}\right)^2 M_{FN} \right]^2,\\
{\cal{M}}_{4} ^2	& = G_{09}\frac{\left( m_eR \right)^2}{8} \left[ \frac{T_1^{(3)}}{g_A} M_{GTN}\right] ^2, \\
{\cal{M}}_{5} ^2	& = G_{09}\frac{\left( m_eR \right)^2}{8} \left[ \frac{F_S^{(3)}g_V}{g_A^2} M_{FN}\right] ^2.
\end{flalign}
\end{subequations}
The parameters $F_S^{(3)}=0.48$ and $T_1^{(3)}=1.38$ are taken form Ref.~\cite{Adler1975}. 
The values of these ${\cal{M}}_{\alpha\beta}^2 $ are presented in Table~\ref{tab-mrond-sr}.
Detailed expressions for $M_{GTN}$ and $M_{FN}$ are presented in the Appendix, and their shell model values are shown in Table~\ref{tab-heavy}.

Considering the $0\nu\beta\beta$ amplitudes displayed in Figs.~\ref{subfig-nonepion} and~\ref{subfig-ntwopion} in the 
one-pion and two-pion exchange modes it is possible to get alternative limits for $\varepsilon_1$ and $\varepsilon_2$ considering a different coefficient, ${\cal{M}}_{\pi N}$.  The analysis of Ref.~\cite{Wodecki1999} suggests these alternative values, here denoted by $\tilde{\varepsilon}_1$ and $\tilde{\varepsilon}_2$, can be obtained as  $\tilde{\varepsilon}_1=\frac{64}{16} \eta_{\pi N}$, and $\tilde{\varepsilon}_2=\frac{2}{3} \eta_{\pi N}$, using

\begin{equation}
 \left[ T^{0\nu}_{1/2}\right]^{-1}=g_{A}^4 \left[ \left| \eta_{\pi N} \right|^2 {\cal{M}}_{\pi N}^2 \right],
\end{equation}
 where
 \begin{flalign}\label{cal_nmes2prime}
\nonumber {\cal{M}}_{\pi N}^2& = G_{01} \left[ c^{1\pi}\left( M_{GT1\pi} + M_{T1\pi} \right) \right. \\ 
		    & + \left. c^{2\pi}\left( M_{GT2\pi} + M_{T2\pi} \right) \right]^2.
 \end{flalign}
The expressions for the factors $c^{1\pi}$ and $c^{2\pi}$ are found in Eq. (151) of Ref.~\cite{Vergados2012}. 
These factors depend on the masses of the up and down quark, and choosing $(m_u+m_d)=11.6$ MeV~\cite{Faessler1998,Horoi2013}, 
one gets $c^{1\pi}=-83.598$, $c^{2\pi}=359.436$ that we use in these calculations. 
The description of $M_\alpha$ (with $\alpha= GT1\pi,\ T1\pi,\ GT2\pi,\ T2\pi$) is presented in the Appendix.
  \begin{table}
\caption{The ${\cal{M}}_{\alpha\beta}^2$ values for the short-range LNV parameters.}
 \begin{tabular}{l@{}lccccc}\label{tab-mrond-sr}
		&				&$^{48}$Ca\ \ \	&$^{76}$Ge\ \ \	&$^{82}$Se\ \ \	&$^{130}$Te\ \ \ &$^{136}$Xe\ \ \ \\ \hline
$10^{13}\cdot$	&${\cal{M}}_{1}^2 $		&$2.63$		&$1.83$		&$6.86$		&$4.83$		&$4.03$	\\
$10^{8}\cdot$	&${\cal{M}}_{2} ^2$		&$0.75$		&$0.54$		&$2.00$		&$1.46$		&$1.21$	 \\
$10^{10}\cdot$	&${\cal{M}}_{3LLz(RRz)}^2$	&$1.30$		&$0.92$		&$3.45$		&$2.50$		&$2.07$ \\
$10^{11}\cdot$	&${\cal{M}}_{3LRz(RLz)}^2$	&$4.82$		&$3.48$		&$13.0$		&$9.54$		&$7.87$ \\
$10^{10}\cdot$	&${\cal{M}}_{4}^2$		&$1.00$		&$0.75$		&$2.68$		&$1.89$		&$1.56$\\
$10^{12}\cdot$	&${\cal{M}}_{5}^2$		&$1.15$		&$0.84$		&$3.01$		&$2.06$		&$1.72$ \\ 
$10^{9}\cdot$	&${\cal{M}}_{\pi N}^2$ 		&$3.36$ 	&$0.87$		&$3.24$		&$2.43$ 	&$1.94$	
\end{tabular}
 \end{table}
 
Shown in Table~\ref{tab-epssr} are the values of the short-range LNV parameters. Using the different hadronization presented in Figs.~\ref{subfig-nonepion} and~\ref{subfig-ntwopion}, 
$\tilde{\varepsilon}_{1}$ provides significantly more stringent upper-limits than $\varepsilon_{1}$. 
With the exception of $^{48}$Ca, where the $\tilde{\varepsilon}_{2}$ limit is identical to $\varepsilon_{2}$, the other $\tilde{\varepsilon}_{2}$ 
upper-limits are almost double those of $\varepsilon_{2}$. Therefore, we conclude that $\varepsilon_{2}$ are better constrained.

\begin{table}
\caption{The ``on-axis`` values of the short-range LNV parameters $\varepsilon^{\beta}_{\alpha}$. The last three lines present the $\eta_{\pi N}$ limits for $\cancel{\cal{R}}_p$ SUSY, and their corresponding $\tilde{\varepsilon}_{1}$ and $\tilde{\varepsilon}_{1}$ limits, respectively.}
 \begin{tabular}{llllll}\label{tab-epssr}
				&$^{48}$Ca		&$^{76}$Ge		&$^{82}$Se		&$^{130}$Te		&$^{136}$Xe \\ \hline
$| \varepsilon_{1} |$		&$8.6\cdot 10^{-6}$	&$2.0\cdot 10^{-7}$	&$1.5\cdot 10^{-6}$	&$4.5\cdot 10^{-7}$	&$9.3\cdot 10^{-8}$	\\
$| \varepsilon_{2} |$		&$5.1\cdot 10^{-8}$	&$1.2\cdot 10^{-9}$	&$8.8\cdot 10^{-9}$	&$2.6\cdot 10^{-9}$	&$5.4\cdot 10^{-10}$	 \\
$| \varepsilon_{3}^{LLz(RRz)}|$	&$3.8\cdot 10^{-7}$	&$8.9\cdot 10^{-9}$	&$6.7\cdot 10^{-8}$	&$2.0\cdot 10^{-8}$	&$4.1\cdot 10^{-9}$ \\
$| \varepsilon_{3}^{LRz(RLz)}|$	&$6.3\cdot 10^{-7}$	&$1.4\cdot 10^{-8}$	&$1.1\cdot 10^{-7}$	&$3.2\cdot 10^{-8}$	&$6.7\cdot 10^{-9}$ \\
$| \varepsilon_{4} |$		&$4.4\cdot 10^{-7}$	&$9.8\cdot 10^{-9}$	&$7.6\cdot 10^{-8}$	&$2.3\cdot 10^{-8}$	&$4.7\cdot 10^{-9}$\\
$| \varepsilon_{5} |$		&$4.1\cdot 10^{-6}$	&$9.3\cdot 10^{-8}$	&$7.1\cdot 10^{-7}$	&$2.2\cdot 10^{-7}$	&$4.5\cdot 10^{-8}$ \\ 
$| \eta_{\pi N} |$		&$7.6\cdot 10^{-8}$	&$2.9\cdot 10^{-9}$	&$2.2\cdot 10^{-8}$	&$6.3\cdot 10^{-9}$	&$1.3\cdot 10^{-9}$ \\
$| \tilde{\varepsilon}_{1} |$	&$3.2\cdot 10^{-7}$	&$1.2\cdot 10^{-8}$	&$9.3\cdot 10^{-8}$	&$2.7\cdot 10^{-8}$	&$5.7\cdot 10^{-9}$ \\
$| \tilde{\varepsilon}_{2} |$	&$5.0\cdot 10^{-8}$	&$1.9\cdot 10^{-9}$	&$1.5\cdot 10^{-8}$	&$4.2\cdot 10^{-9}$	&$8.9\cdot 10^{-10}$	
\end{tabular}
 \end{table}
 
\section{Discussions} \label{discussions}
From the $\eta_{0\nu}$ limits presented in Table~\ref{tab-mrond-light} for $^{136}$Xe, one gets the lowest shell model upper-limit for the Majorana neutrino mass $\left< m_{\beta\beta}\right>\sim 140$ meV. A wider range of values, $60-165$ meV can be found if the NME calculated with a larger number of nuclear models are considered~\cite{kamlandzen16}.

Considering the diagram in Fig.~\ref{subfig-npion_nu}, it is possible to get lower limits for $\epsilon^{TR}_{TR}$, denoted as $\tilde{\epsilon}^{TR}_{TR}$ in Table~\ref{tab-epslr}, than those corresponding to the diagram in Fig.~\ref{subfig-nlongrange}, with the exception of $^{48}$Ca, as can be seen in Table~\ref{tab-epslr}.
Considering the different hadronization scenario presented in Figs.~\ref{subfig-nonepion} and~\ref{subfig-ntwopion}, $\tilde{\varepsilon}_{1}$ provides a significantly more stringent upper-limits than $\varepsilon_{1}$. 
With the exception of $^{48}$Ca, where the $\tilde{\varepsilon}_{2}$ limit is identical to $\varepsilon_{2}$, the other $\tilde{\varepsilon}_{2}$ upper-limits are almost double those of $\varepsilon_{2}$.
\begin{table}
\caption{The  BSM effective scale (in GeV) for different dimension-D operators at the present $^{136}$Xe half-life limit ($\Lambda^0_D$) and for $T_{1/2}\approx1.1\times 10^{28}$ years ($\Lambda_D$).}
 \begin{tabular}{@{}lccrrr@{}}\label{tab-operator-limits}
${\cal{O}}_D$	 & $\ \ \bar{\epsilon}_D$ 	& $\ \ \Lambda^0_D(y=1)$ 	& $\ \ \Lambda^0_D(y=y_e)$ & $\ \ \Lambda_D(y=y_e)$ \\ \hline
${\cal{O}}_5$	 & $\ \ 2.8 \cdot 10^{-7}$	&$\ \ 2.12 \cdot 10^{14}$	& 1904 & 19044		\\
${\cal{O}}_7$	 &$\ 2.0 \cdot 10^{-7}$		&$\ \ 3.76 \cdot 10^{4\ }$	& 542	& 1169		 \\
${\cal{O}}_9$	 &$\ \ 9.3 \cdot 10^{-8}$	&$\ \ 2.72 \cdot 10^{3\ }$	& 2718 & 4307	\\
${\cal{O}}_{11}$ &$\ \ 9.3 \cdot 10^{-8}$	&$\ \ 1.24 \cdot 10^{3\ }$	& 33	& 46 	 \\
\end{tabular}
 \end{table}
 
As suggested in Ref.~\cite{Deppisch2015-PRD92} (see the diagrams of their Fig.1), at the electroweak scale when the appropriate Higgs fields are included, the diagram 1.b originates from a dimension-5 BSM Lagrangian, $\mathcal{O}_5$, responsible for the Majorana neutrino mass. Similarly the low-energy dimension-6 Lagrangian $\mathcal{L}_6$ corresponds to a dimension-7 BSM operator, $\mathcal{O}_7$, and the low energy dimension-9 Lagrangian $\mathcal{L}_9$ can be rearranged as dimension-9 and dimension-11 operators, $\mathcal{O}_9$ and $\mathcal{O}_{11}$. Using the effective field theory one can infer the energy scale $\Lambda_D$ up to which these effective field operators are not broken:

\begin{equation}
\mathcal{L}_D=\frac{g}{\left(\Lambda_D\right)^{D-4}} \mathcal{O}_D ,
\label{l-ews}
\end{equation}
where $D$ is the dimension of the effective field operator. Here $g$ is considered to be a dimensionless coupling constant of the order of 1.
Following Ref.~\cite{Deppisch2015-PRD92} one can find relations between the constants entering our $\mathcal{L}_6$ and $\mathcal{L}_9$ Lagrangian and the effective field theory Lagrangians above the electroweak scale, Eq. (\ref{l-ews}).

\begin{align}
&m_e\bar{\epsilon}_5=\frac{g^2(yv)^2}{\Lambda_5}, & & \frac{G_F \bar{\epsilon}_7}{\sqrt{2}}=\frac{g^3(yv)}{2(\Lambda_7)^3},& \nonumber \\
&\frac{G^2_F \bar{\epsilon}_9}{2m_p}=\frac{g^4}{(\Lambda_9)^5}, & & \frac{G^2_F \bar{\epsilon}_{11}}{2m_p}=\frac{g^6(yv)^2}{(\Lambda_{11})^7}.&
\end{align}
Here, $m_e=0.511\times 10^{-3}$ GeV is the electron mass, $g=1$ is a generic coupling constant, $v=174$ GeV is the Higgs vacuum expectation value, $y$ is a Yukawa coupling associated to the interaction with the Higgs bosons,
$G_F= 1.166\times 10^{-5}$ GeV$^{-2}$ is the Fermi coupling constant, and $m_p=0.938$ GeV is the proton mass.  The $\bar{\epsilon}_D$ (with $D=\{ 5,\ 7,\ 9,\ 11 \}$) can be extracted from the LNV parameters in Eqs. (\ref{lag-longrange}) and (\ref{lag9}). Considering that values of these LNV parameters may be affected by mixing angles that might distort the scales in Eq. (\ref{l-ews}), we choose their maximum values: 
$\bar{\epsilon}_5=|\eta_{0\nu}|$, $\bar{\epsilon}_7=\textmd{Max}\Bigl[ |\epsilon^{V+A}_{V-A}| \Bigr.$, $|\epsilon^{V+A}_{V+A}|$, 
$|\epsilon^{S+P}_{S\pm P}|$, $|\epsilon^{TR}_{TL}|$, $\Bigl. |\epsilon^{TR}_{TR}| \Bigr]$, 
$\bar{\epsilon}_9=\textmd{Max}\Bigl[ |\varepsilon_1| \Bigr.$, $|\varepsilon_{2}|$, $|\varepsilon_{3}^{LLz(RRz)}|$, 
$|\varepsilon_{3}^{LRz(RLz)}|$, $|\varepsilon_{4}|$, $\Bigl. |\varepsilon_5| \Bigr]$, and $\bar{\epsilon}_{11}=\bar{\epsilon}_{9}$.

To extract the limits of the BSM scales $\Lambda_{5,7,9,11}$ we need the most stringent limits for the LNV parameters, which are found for the case of $^{136}$Xe. 
Inspecting Tables~\ref{tab-epslr} and~\ref{tab-epssr} we found that $\bar{\epsilon}_5$ corresponds to the $\eta_{0\nu}$ parameter of the light left-handed Majorana neutrino exchange mechanism. For $\bar{\epsilon}_7$ we choose $\epsilon^{V+A}_{V+A}$, that is the largest long-range $\epsilon_\alpha^\beta$ parameter. In the case of $\bar{\epsilon}_{9}=\bar{\epsilon}_{11}$ we select $\varepsilon_1$, being the largest short-range $\varepsilon_\alpha^\beta$ parameter. These values are listed in Table~\ref{tab-operator-limits}.

As in Ref.~\cite{Deppisch2015-PRD92} we take $g=1$ in Eq. (\ref{l-ews}). However, we introduce here the Yukawa coupling $y$ between the Higgs boson field and the fermion fields, and we consider two cases: (i) $y=1$ corresponding to the top quark mass (choice made in Ref.~\cite{Deppisch2015-PRD92}), and (ii) $y=3\times 10^{-6}$ corresponding to the electron mass. Based on these values we calculate the limits of the new BSM scales or different dimension-D operators. The results are shown in Table~\ref{tab-operator-limits}. The $\Lambda^0_D$ scales are calculated using the present lower limit for the half-life of $^{136}$Xe, $1.1\times 10^{26}$. $\Lambda_D$ is estimated assuming a half-life of $T_{1/2}\approx1.1\times 10^{28}$ years, which would correspond to a $\left< m_{\beta\beta}\right>\approx 14$ meV.

The $\Lambda_9$ scale does not depend on the unknown Yukawa coupling, and from that point of view, if $\mathcal{O}_9$ amplitude is dominant, that would indicate that the scale of new physics should be found around 3 TeV. Unfortunately, the $\Lambda_9$ scale, as well as all other high $D$ scales, are not very sensitive to the $0\nu\beta\beta$ half-life, because they scale as $T_{1/2}^{\frac{1}{2(D-4)}}$. $\mathcal{O}_7$ and $\mathcal{O}_{11}$ provide small low-limits for $\Lambda_7$ and $\Lambda_{11}$. This feature is likely related to the fact that these terms are originating from small term in the mixing matrix (e.g. the small $S$ matrix in Eq. (A3) of~\cite{HoroiNeacsu2016prd}), and thus $g \sim 1$ in Eq. (\ref{l-ews}) is not a good choice. The most sensitive scale to both the unknown Yukawa and the $0\nu\beta\beta$ half-life is $\Lambda_5$. Assuming a Yukawa coupling corresponding to the electron mass, one can conclude that the $0\nu\beta\beta$ decay could be consistent with a new physics scale somewhere between 2 TeV and 20 TeV.

\section{Conclusions} \label{section-conclusions}
This work advances and extends the analysis of 
BSM physics parameters involved in the neutrinoless double-beta decay. 
We calculate 23 nuclear matrix elements and 9 phase-space factors. Five of these nuclear matrix elements ($M_{GT^\prime}$, $M_{GT^{\prime\prime}}$, $M_{F^\prime}$, $M_{T^\prime}$, and $M_{T^{\prime\prime}}$) are calculated for the first time using shell model techniques. Three new hadron-level diagrams, Fig. 2.e, 2.f, 2.g are for the first time considered in the full analyses based on the effective field theory approach to $0\nu\beta\beta$ decay (they were only considered in the past in the context of particular mechanisms).

Using a general effective field theory and assuming that one LNV coupling plays a dominant contribution to the $0\nu\beta\beta$ decay amplitude, we extract limits for the effective Majorana mass and 11 effective low-energy couplings in the case of five nuclei of immediate experimental interest. 
Due to the better half-life limits, the most stringent limits for the LNV couplings are found for $^{136}$Xe, closely followed by $^{76}$Ge. 
An upper-limit for the Majorana neutrino mass $\left< m_{\beta\beta}\right>$ of 140 meV was calculated in the case of $^{136}$Xe. Assuming a Yukawa coupling corresponding to the electron mass, one can conclude that the $0\nu\beta\beta$ decay could be consistent with a new physics scale somewhere between 2 TeV and 20 TeV.

Using the upper limits for the LNV coupling we extract limits for the energy scale of the new physics, using EFT arguments. We found that the scale associated with the dimension-9 EFT operator is stable, and indicates a new physics scale around 3 TeV. We also found that the dimension-5 EFT operator associated with the Majorana neutrino mass varies significantly with the Yukawa coupling to Higgs and the $0\nu\beta\beta$ decay half-life. 

Should neutrinoless double-beta decay be experimentally observed, a thorough analysis of the outgoing electrons angular and energy distributions (presented in Ref.~\cite{HoroiNeacsu2016prd}) based on accurate calculations of the nuclear matrix elements is needed to investigate subsets of these LNV couplings and identify the presence of the right-handed currents.

\section{Appendix} \label{section-appendix}
In this Appendix, we present the detailed expressions for the ${\cal{M}}_{i}^2$ coefficients that are needed to analyze the outcome of Eq.~ (\ref{flifetime}).

The NME that enter the equations (\ref{cal_nmes},~\ref{cal_nmes1},~\ref{cal_nmes1prime},~\ref{cal_nmes2}, and~\ref{cal_nmes2prime}) are written as a product of two-body transition densities (TBTD) and two-body matrix elements (TBME), where the summation is over all the nucleon states. Their numerical values when calculated within the shell model
approach are presented in Table~\ref{tab-nme-light} for the light left-handed Majorana neutrino exchange, in Table~\ref{tab-nme-muto} for the long-range part in Fig.~\ref{fig-ndiagrams}, and in Table~\ref{tab-heavy} for the short-range component of Fig.~\ref{fig-ndiagrams}. The general expressions for the NME are (see Refs.~\cite{Doi1985,Horoi2013,HoroiNeacsu2016prd}):
\begin{flalign}\label{eq-Ma}
\nonumber M_{\alpha}&=\sum_{j_p j_{p'} j_n j_{n'} J^\pi} TBTD\left( j_p j_{p'},j_n j_{n'};J^\pi \right) \\
&\times \left< j_p j_{p'}; J^\pi \left\| \tau_{-1} \tau_{-2} {\cal{O}}^{\gamma,\phi,\theta,P,R}_{12} \right\| j_n j_{n'}; J^\pi \right> . 
\end{flalign} 
We group the operators that share similar structure into five families.
\begin{flalign}
&\nonumber \textmd{Gamow-Teller operator}: {\cal{O}}^{\gamma}_{12}  =  \vv{\sigma}_1 \cdot \vv{\sigma}_2 H_{\gamma}(r),   \\
&\nonumber \textmd{Fermi operator}:  {\cal{O}}^{\phi}_{12} \ =  H_{\phi}(r),  \\
&\nonumber \textmd{Tensor operator}:  {\cal{O}}^{\theta}_{12} =  \left[3(\vv{\sigma}_1 \cdot {\bf \hat{r}})(\vv{\sigma}_2 \cdot {\bf \hat{r}}) - \vv{\sigma}_1 \cdot \vv{\sigma}_2 \right] H_{\theta}(r),\\
&\nonumber \textmd{P operator}: {\cal{O}}^{P}_{12} =  \left( \vv{\sigma}_1 - \vv{\sigma}_2 \right) H_{P}(r),\\
&\nonumber \textmd{R operator}: {\cal{O}}^{R}_{12} =  \vv{\sigma}_1 \cdot \vv{\sigma}_2 H_{R}(r).
\end{flalign}
Here, $\gamma=$ $GT$, $GT\omega$, $GTq$, $GTN$, $GT'$, $GT''$, $GT\pi\nu$, $GT1\pi$, $GT2\pi$, 
$\phi=F,\ F\omega,\ Fq,\ FN,\ F'$, and $\theta=T,\ Tq,\ T',\ T'',\ T\pi\nu,\ T1\pi,\ T2\pi$.
Equations (\ref{eq-qnmes}) present the radial part of the NME and their expressions are adapted for consistency from 
Refs.~\cite{Doi1985},\cite{Deppisch2012}, and~\cite{Vergados2012}.
 \begin{table}
\caption{NME values for the exchange of light left-handed Majorana neutrinos corresponding to the diagram in Fig.~\ref{subfig-nlight}.}
 \begin{tabular}{lrrrrr} \label{tab-nme-light}
		&$^{48}$Ca	&$^{76}$Ge&$^{82}$Se&$^{130}$Te	&$^{136}$Xe \\ \hline
$M_{GT}$	&$ 0.807$	&$ 3.206$ &$ 3.005$ &$ 1.662$	&$ 1.505$ \\
$M_{F}$		&$-0.233$	&$-0.674$ &$-0.632$ &$-0.438$	&$-0.400$ \\
$M_{T}$		&$ 0.080$	&$ 0.011$ &$ 0.012$ &$-0.007$	&$-0.008$ 
\end{tabular}
\end{table}
\begin{table}
\caption{NME for the long-range part shown in Figs.~\ref{subfig-nlongrange} and~\ref{subfig-npion_nu}.}
 \begin{tabular}{lrrrrr} \label{tab-nme-muto}
		&$^{48}$Ca	&$^{76}$Ge&$^{82}$Se&$^{130}$Te	&$^{136}$Xe \\ \hline
$M_{GTq}$	&$ 0.709$	&$ 3.228$ &$ 3.034$ &$ 1.587$	&$ 1.440$ \\
$M_{GT\omega}$	&$ 0.930$	&$ 3.501$ &$ 3.287$ &$ 1.855$	&$ 1.682$ \\
$M_{GT^\prime}$	&$ 0.841$	&$ 2.699$ &$ 2.567$ &$ 2.120$	&$ 1.935$ \\
$M_{GT^{\prime\prime}}$	&$ 3.581$	&$11.982$ &$11.490$ &$12.210$	&$11.202$ \\
$M_{GT\pi\nu}$	&$ 86.2$	&$331.6$  &$ 313.3$ &$ 184.2$	&$167.7$ \\
$M_{Fq}$	&$-0.121$	&$-0.383$ &$-0.362$ &$-0.249$	&$-0.230$ \\
$M_{F \omega}$	&$-0.232$	&$-0.659$ &$-0.618$ &$-0.427$	&$-0.391$ \\
$M_{F^\prime}$	&$-0.258$	&$-0.812$ &$-0.772$ &$-0.635$	&$-0.581$ \\
$M_{Tq}$	&$-0.173$	&$-0.059$ &$-0.058$ &$-0.013$	&$-0.012$ \\
$M_{T^\prime}$	&$ 0.337$	&$ 0.015$ &$ 0.025$ &$-0.077$	&$-0.085$ \\
$M_{T^{\prime\prime}}$	&$ 2.231$	&$ 0.028$ &$0.118$ &$-0.773$	&$-0.861$ \\
$M_{T\pi\nu}$&$ 21.3$	&$  7.3$  &$   6.9$ &$   1.2$	&$  1.1$ \\ 
$M_{P}$		&$ 0.395$	&$-2.466$ &$-2.332$ &$-1.729$	&$-1.617$ \\
$M_{R}$		&$ 1.014$	&$ 3.284$ &$ 3.127$ &$ 2.562$	&$ 2.341$ 
\end{tabular}
 \end{table}
  \begin{table}
\caption{The short-range NME involved in Figs.~\ref{subfig-nheavy},~\ref{subfig-nonepion}, and~\ref{subfig-ntwopion}.}
 \begin{tabular}{lrrrrr}\label{tab-heavy}
		&$^{48}$Ca	&$^{76}$Ge&$^{82}$Se&$^{130}$Te	&$^{136}$Xe \\ \hline
$M_{GTN}$	&$ 58.5$	&$162.3$  &$ 150.1$ &$ 107.6$	&$ 96.6$ \\
$M_{GT 1\pi }$	&$-1.354$ 	&$-3.559$ &$-3.282$ &$-2.421$	&$-2.171$ \\
$M_{GT 2\pi}$	&$-0.676$	&$-1.983$ &$-1.854$ &$-1.257$	&$-1.136$ \\
$M_{FN}$	&$-22.9$	&$-62.6$  &$ -58.1$ &$ -41.0$	&$-36.9$ \\ 
$M_{T 1\pi}$ 	&$-0.590$	&$-0.010$ &$-0.027$ &$0.106$	&$0.115$ \\ 
$M_{T 2\pi}$ 	&$-0.227$	&$-0.010$ &$-0.015$ &$0.038$	&$0.040$ 
\end{tabular}
 \end{table}
\begin{subequations}
 \begin{align}
H_{GT}&=\frac{2R}{\pi}\int \frac{h^2_{GT}(q^2)}{q(q+\bar{E})} j_0(qr) q^2 \dd{q},  \\
H_{GT\omega}&=\frac{2R}{\pi}\int \frac{h_{A}^2(q^2)}{(q+\bar{E})^2} j_0(qr) q^2 \dd{q},  \\
H_{GTq}&=\frac{2R}{\pi}r\int \frac{h_{A}^2(q^2)}{q+\bar{E}} j_1(qr) q^2 \dd{q},  \\
H_{GTN}&=\frac{2R}{\pi m_e m_p}\int h_{A}^2(q^2)j_0(qr) q^2 \dd{q},  \\
H_{GT'}&=\frac{2R^2}{\pi m_p}\int \frac{q^2 h_{A}^2(q^2)}{q(q+\bar{E})} j_0(qr) q^2 \dd{q},  \\
H_{GT''}&=\frac{2R^3}{\pi m_p}\int \frac{q^2 h_{A}^2(q^2)}{q+\bar{E}} j_0(qr) q^2 \dd{q},  \\
H_{GT\pi\nu}&=\frac{2R}{\pi}\int \frac{h^2_{GT\pi\nu}(q^2)}{q(q+\bar{E})} j_0(qr) q^2 \dd{q},  \\
H_{GT1\pi}&=-\frac{2R}{\pi}\int h_{A}^2(q^2)\frac{q^2/m_\pi^4}{1+q^2/m_\pi^2} j_0(qr) q^2 \dd{q},  \\
H_{GT2\pi}&= -\frac{4R}{\pi}\! \! \int \! \! h_{A}^2(q^2)\frac{q^2/m_\pi^4}{(1+q^2/m_\pi^2)^2} j_0(qr) q^2 \dd{q},  \\
H_{F}&=\frac{2R}{\pi}\int \frac{h_{V}^2(q^2)}{q(q+\bar{E})} j_0(qr) q^2 \dd{q},  \\
H_{F\omega}&=\frac{2R}{\pi}\int \frac{h_{V}^2(q^2)}{(q+\bar{E})^2} j_0(qr) q^2 \dd{q},  \\
H_{Fq}&=\frac{2R}{\pi}r\int \frac{h_{V}^2(q^2)}{q+\bar{E}} j_1(qr) q^2 \dd{q},  \\
H_{FN} &=\frac{2R}{\pi m_e m_p}\int h_{V}^2(q^2)j_0(qr) q^2 \dd{q},  \\
H_{F'} &=\frac{2R^2}{\pi m_p}\int \frac{q^2 h_{V}^2(q^2)}{q(q+\bar{E})} j_0(qr) q^2 \dd{q},  \\
H_{T}&=- \frac{2R}{\pi}\int \frac{h^2_{T}(q^2)}{q(q+\bar{E})} j_2(qr) q^2 \dd{q},  \\
H_{Tq}&=\frac{2R}{3\pi}\sqrt{\frac{2}{3}}r C^{(2)}({\bf \hat{r}})\int \frac{h_{A}^2(q^2)}{q+\bar{E}} j_1(qr) q^2 \dd{q},  \label{eq-tq} \\
H_{T'}&=-\frac{2R^2}{\pi m_p}\int \frac{q^2 h_{A}^2(q^2)}{q(q+\bar{E})} j_2(qr) q^2 \dd{q},  \\
H_{T''}&=-\frac{2R^3}{\pi m_p}\int \frac{q^2 h_{A}^2(q^2)}{q+\bar{E}} j_2(qr) q^2 \dd{q},  \\
H_{T\pi\nu}&=- \frac{2R}{\pi}\int \frac{h^2_{T\pi\nu}(q^2)}{q(q+\bar{E})} j_2(qr) q^2 \dd{q},  \\
H_{T1\pi} &= \frac{2R}{\pi}\int h_{A}^2(q^2) \frac{q^2/m_\pi^4}{1+q^2/m_\pi^2} j_2(qr) q^2 \dd{q},  \\
H_{T2\pi} &= \frac{4R}{\pi}\! \int \! h_{A}^2(q^2) \frac{q^2/m_\pi^4}{(1+q^2/m_\pi^2)^2} j_2(qr) q^2 \dd{q},  \\
\nonumber H_{R}&=\frac{(\mu_p - \mu_n)}{3} \frac{g_V}{g_A}\frac{2R^2}{\pi m_p} \\
	        &\times \int q \frac{h_{A}(q^2)h_{V}(q^2)}{q+\bar{E}} j_0(qr) q^2 \dd{q},  \\
\nonumber H_{P}&=\sqrt{2}\frac{2R}{\pi}\frac{g_V}{g_A}C^{(1)}({\bf \hat{r}}) C^{(1)}({\bf \hat{r}_+}) r_+ \\
	      & \times \int \frac{h_{A}(q^2)h_{V}(q^2)}{q+\bar{E}} j_1(qr) q^2 \dd{q},  \label{eq-pq}.
\end{align} \label{eq-qnmes}
\end{subequations}
Here, the expressions of $C^{(L)}_M$ and $\bf r$ of Eqs.~(\ref{eq-tq},~\ref{eq-pq}) are: 
\begin{equation*}
 C^{(L)}_M=\sqrt{\frac{4\pi}{2L+1}}Y_{LM},
\end{equation*}
\begin{align*}
 &{\bf r} ={\bf r_1} - {\bf r_2},& &{\bf r_+}=\frac{{\bf r_1}+{\bf r_2}}{2},&  &r=|{\bf r}|, \\ 
 &{\bf \hat{r}}=\frac{\bf r}{r},&  &r_+=|{\bf r_+}|,& &{\bf \hat{r}_+}=\frac{\bf r_+}{r_+}.
\end{align*}
%
The finite-size effects are taken into account via the following dipole form-factors:
\begin{subequations}
\begin{align}
g_A(q^2)=&\left( \frac{\lambda_A^2}{\lambda_A^2+q^2} \right) ^2, \\
g_V(q^2)=&\left( \frac{\lambda_V^2}{\lambda_V^2+q^2} \right) ^2, \\
g_M(q^2)=&\left(\mu_p - \mu_n\right)g_V(q^2).
\end{align}
\end{subequations}
Here $\lambda_A=1086$ MeV and $\lambda_V=850$ MeV are the axial and vector momentum cutoffs, respectively, and $(\mu_p-\mu_n)\simeq 3.7$.

The form-factors entering Eqs.~\ref{eq-qnmes} are:
\begin{subequations}
\begin{align}
h_{V}(q^2) =& g_V(q^2), \\
h_{A}(q^2) =& g_A(q^2), \\
\nonumber h^2_{GT}(q^2)=& g_A^2(q^2)\left[ 1- \frac{2}{3} \frac{q^2}{q^2+{m_\pi}^2} + \frac{1}{3} \left( \frac{q^2}{q^2+m_\pi^2} \right)^2 \right] \\
	     +& \frac{2}{3} \frac{g_M^2(q^2)}{g_A^2} \frac{q^2}{4m_p^2}, \\
\nonumber h^2_{T}(q^2)=& g_A^2(q^2)\left[ \frac{2}{3} \frac{q^2}{q^2+{m_\pi}^2} - \frac{1}{3} \left( \frac{q^2}{q^2+m_\pi^2} \right)^2 \right] \\
	     +& \frac{1}{3} \frac{g_M^2(q^2)}{g_A^2} \frac{q^2}{4m_p^2}, \\
h^2_{GT\pi\nu}(q^2)=& - \frac{g^2_A(q^2)}{6}\frac{m^4_\pi}{m_e(m_u+m_d)}\frac{q^2}{(q^2+m^2_\pi)^2}, \! \! \\
h^2_{T\pi\nu}(q^2)=& h^2_{GT\pi\nu}(q^2).
\end{align}
\end{subequations}
$m_e=0.511$ MeV is the electron mass, $m_\pi=139$ MeV is the pion mass, $m_p=938$ MeV is the proton mass, and the quark masses sum is $(m_u+m_d)=11.6$ MeV~\cite{Faessler1998,Horoi2013}. 

The NME presented in this section (Eq.~(\ref{eq-Ma})) are calculated using shell model approaches. To take into account the two-nucleon short-range correlation (SRC) 
we multiply the the relative wave functions by $f(r) = 1-ce^{-ar^2} (1-br^2 )$; in the CD-Bonn parametrization used here 
$a=1.52$ fm$^{-2}$, $b=1.88$ fm$^{-2}$, and $c=0.46$ fm$^{-2}$~\cite{Simkovic2009}. This method is described in greater detail in 
Refs.~\cite{HoroiStoicaBrown2007,HoroiStoica2010,NeacsuStoicaHoroi2012,HoroiBrown2013,SenkovHoroi2013,Horoi2013,BrownHoroiSenkov2014,SenkovHoroi2014,SenkovHoroiBrown2014,
NeacsuHoroi2015,NeacsuHoroi2016}.
The signs of all the NME presented in the following tables are relative to the sign of $M_{GT}$, which is taken to be positive.
Table~\ref{tab-nme-light} presents the  $M_{GT}$, $M_F$, and $M_T$  NME involved in the standard mass mechanism with left-handed currents of Eq. (\ref{cal_nmes}).
For these NME, an optimal closure energy $\left< \bar{E} \right>$ was used for each effective Hamiltonian~\cite{SenkovHoroiBrown2014}:
$\left< \bar{E} \right>=0.5$MeV for $^{48}$Ca~\cite{SenkovHoroi2013} and the GXPF1A Hamiltonian~\cite{Honma2005}, 
$\left< \bar{E} \right>=3.4$MeV for $^{76}$Ge~\cite{Senkov2016} and $^{82}$Se~\cite{SenkovHoroiBrown2014} calculated with the JUN45 Hamiltonian~\cite{JUN45}, 
and $\left< \bar{E} \right>=3.5$MeV for $^{130}$Te~\cite{NeacsuHoroi2015} and $^{136}$Xe~\cite{HoroiBrown2013} calculated with the SVD Hamiltonian~\cite{Chong2012}.

 The long-range NME $M_\alpha$ (with $\alpha=$ $GTq$, $Fq$, $Tq$, $GT\omega$, $F\omega$, $P$, $R$, $GT'$, $F'$, $T'$, $GT''$, $T''$) that appear in 
 Eq. (\ref{cal_nmes1}) and $M_{GT{\pi\nu}}$ and $M_{T{\pi\nu}}$ of Eq. (\ref{cal_nmes1prime}) are presented in Table~\ref{tab-nme-muto}.

Shown in Table~\ref{tab-heavy} are the short-range NME $M_{GTN}$ and $M_{FN}$ that appear in Eq. (\ref{cal_nmes2} and 
$M_\alpha$ (with $\alpha=$ $GT1\pi$, $T1\pi$, $GT2\pi$, $T2\pi$) in Eq.~\ref{cal_nmes2prime}).

\section{ACKNOWLEDGMENTS}
Support from  the NUCLEI SciDAC Collaboration under U.S. Department of Energy Grant No. DE-SC0008529  is acknowledged. M. Horoi also acknowledges the U.S. NSF Grant No. PHY-1404442 and the U.S. Department of Energy Grant No. DE-SC0015376.
\bibliographystyle{apsrev}%
\bibliography{bb}%

\begin{thebibliography}{90}
\expandafter\ifx\csname natexlab\endcsname\relax\def\natexlab#1{#1}\fi
\expandafter\ifx\csname bibnamefont\endcsname\relax
  \def\bibnamefont#1{#1}\fi
\expandafter\ifx\csname bibfnamefont\endcsname\relax
  \def\bibfnamefont#1{#1}\fi
\expandafter\ifx\csname citenamefont\endcsname\relax
  \def\citenamefont#1{#1}\fi
\expandafter\ifx\csname url\endcsname\relax
  \def\url#1{\texttt{#1}}\fi
\expandafter\ifx\csname urlprefix\endcsname\relax\def\urlprefix{URL }\fi
\providecommand{\bibinfo}[2]{#2}
\providecommand{\eprint}[2][]{\url{#2}}

\bibitem[{\citenamefont{Schechter and Valle}(1982)}]{SchechterValle1982}
\bibinfo{author}{\bibfnamefont{J.}~\bibnamefont{Schechter}} \bibnamefont{and}
  \bibinfo{author}{\bibfnamefont{J.~W.~F.} \bibnamefont{Valle}},
  \bibinfo{journal}{Phys. Rev. D} \textbf{\bibinfo{volume}{25}},
  \bibinfo{pages}{2951} (\bibinfo{year}{1982}).

\bibitem[{\citenamefont{Nieves}(1984)}]{Nieves1984}
\bibinfo{author}{\bibfnamefont{J.}~\bibnamefont{Nieves}},
  \bibinfo{journal}{Phys. Lett. B} \textbf{\bibinfo{volume}{147}},
  \bibinfo{pages}{375} (\bibinfo{year}{1984}).

\bibitem[{\citenamefont{Takasugi}(1984)}]{Takasugi1984}
\bibinfo{author}{\bibfnamefont{E.}~\bibnamefont{Takasugi}},
  \bibinfo{journal}{Phys. Lett. B} \textbf{\bibinfo{volume}{149}},
  \bibinfo{pages}{372} (\bibinfo{year}{1984}).

\bibitem[{\citenamefont{Hirsch et~al.}(2006)\citenamefont{Hirsch, Kovalenko,
  and Schmidt}}]{Hirsch2006}
\bibinfo{author}{\bibfnamefont{M.}~\bibnamefont{Hirsch}},
  \bibinfo{author}{\bibfnamefont{S.}~\bibnamefont{Kovalenko}},
  \bibnamefont{and} \bibinfo{author}{\bibfnamefont{I.}~\bibnamefont{Schmidt}},
  \bibinfo{journal}{Phys. Lett. B} \textbf{\bibinfo{volume}{642}},
  \bibinfo{pages}{106} (\bibinfo{year}{2006}).

\bibitem[{\citenamefont{Pati and Salam}(1974)}]{PatiSalam1974}
\bibinfo{author}{\bibfnamefont{J.}~\bibnamefont{Pati}} \bibnamefont{and}
  \bibinfo{author}{\bibfnamefont{A.}~\bibnamefont{Salam}},
  \bibinfo{journal}{Phys. Rev. D} \textbf{\bibinfo{volume}{10}},
  \bibinfo{pages}{275} (\bibinfo{year}{1974}).

\bibitem[{\citenamefont{Mohapatra and
  Pati}(1975{\natexlab{a}})}]{MohapatraPati1975}
\bibinfo{author}{\bibfnamefont{R.}~\bibnamefont{Mohapatra}} \bibnamefont{and}
  \bibinfo{author}{\bibfnamefont{J.}~\bibnamefont{Pati}},
  \bibinfo{journal}{Phys. Rev. D} \textbf{\bibinfo{volume}{11}},
  \bibinfo{pages}{2558} (\bibinfo{year}{1975}{\natexlab{a}}).

\bibitem[{\citenamefont{Senjanovic and Mohapatra}(1975)}]{Senjanovic1975}
\bibinfo{author}{\bibfnamefont{G.}~\bibnamefont{Senjanovic}} \bibnamefont{and}
  \bibinfo{author}{\bibfnamefont{R.~N.} \bibnamefont{Mohapatra}},
  \bibinfo{journal}{Phys. Rev. D} \textbf{\bibinfo{volume}{12}},
  \bibinfo{pages}{1502} (\bibinfo{year}{1975}).

\bibitem[{\citenamefont{Keung and Senjanovic}(1983)}]{KeungSenjanovic1983}
\bibinfo{author}{\bibfnamefont{W.-Y.} \bibnamefont{Keung}} \bibnamefont{and}
  \bibinfo{author}{\bibfnamefont{G.}~\bibnamefont{Senjanovic}},
  \bibinfo{journal}{Phys. Rev. Lett.} \textbf{\bibinfo{volume}{50}},
  \bibinfo{pages}{1427} (\bibinfo{year}{1983}).

\bibitem[{\citenamefont{Barry and Rodejohann}(2013)}]{Barry2013}
\bibinfo{author}{\bibfnamefont{J.}~\bibnamefont{Barry}} \bibnamefont{and}
  \bibinfo{author}{\bibfnamefont{W.}~\bibnamefont{Rodejohann}},
  \bibinfo{journal}{J. High Energy Phys.} p. \bibinfo{pages}{153}
  (\bibinfo{year}{2013}).

\bibitem[{\citenamefont{Khachatryan et~al.}(2014)\citenamefont{Khachatryan,
  Sirunyan, Tumasyan, Adam, Bergauer, Dragicevic, Erö, Fabjan, Friedl,
  Fruhwirth et~al.}}]{CMS2014}
\bibinfo{author}{\bibfnamefont{V.}~\bibnamefont{Khachatryan}},
  \bibinfo{author}{\bibfnamefont{A.~M.} \bibnamefont{Sirunyan}},
  \bibinfo{author}{\bibfnamefont{A.}~\bibnamefont{Tumasyan}},
  \bibinfo{author}{\bibfnamefont{W.}~\bibnamefont{Adam}},
  \bibinfo{author}{\bibfnamefont{T.}~\bibnamefont{Bergauer}},
  \bibinfo{author}{\bibfnamefont{M.}~\bibnamefont{Dragicevic}},
  \bibinfo{author}{\bibfnamefont{J.}~\bibnamefont{Erö}},
  \bibinfo{author}{\bibfnamefont{C.}~\bibnamefont{Fabjan}},
  \bibinfo{author}{\bibfnamefont{M.}~\bibnamefont{Friedl}},
  \bibinfo{author}{\bibfnamefont{R.}~\bibnamefont{Fruhwirth}},
  \bibnamefont{et~al.} (\bibinfo{collaboration}{CMS-Collaboration}),
  \bibinfo{journal}{Eur. Phys. J. C} \textbf{\bibinfo{volume}{74}},
  \bibinfo{pages}{3149} (\bibinfo{year}{2014}).

\bibitem[{\citenamefont{Horoi and
  Neacsu}(2016{\natexlab{a}})}]{HoroiNeacsu2016prd}
\bibinfo{author}{\bibfnamefont{M.}~\bibnamefont{Horoi}} \bibnamefont{and}
  \bibinfo{author}{\bibfnamefont{A.}~\bibnamefont{Neacsu}},
  \bibinfo{journal}{Phys. Rev. D} \textbf{\bibinfo{volume}{93}},
  \bibinfo{pages}{113014} (\bibinfo{year}{2016}{\natexlab{a}}),
  \eprint{arXiv:1511.00670 [hep-ph]}.

\bibitem[{\citenamefont{Neacsu and Horoi}(2016)}]{Neacsu2016ahep-dist}
\bibinfo{author}{\bibfnamefont{A.}~\bibnamefont{Neacsu}} \bibnamefont{and}
  \bibinfo{author}{\bibfnamefont{M.}~\bibnamefont{Horoi}},
  \bibinfo{journal}{Advances in High Energy Physics}
  \textbf{\bibinfo{volume}{2016}} (\bibinfo{year}{2016}).

\bibitem[{\citenamefont{Cirigliano
  et~al.}(2017{\natexlab{a}})\citenamefont{Cirigliano, Dekens, de~Vries,
  Graesser, and Mereghetti}}]{Cirigliano2017-arxiv}
\bibinfo{author}{\bibfnamefont{V.}~\bibnamefont{Cirigliano}},
  \bibinfo{author}{\bibfnamefont{W.}~\bibnamefont{Dekens}},
  \bibinfo{author}{\bibfnamefont{J.}~\bibnamefont{de~Vries}},
  \bibinfo{author}{\bibfnamefont{M.~L.} \bibnamefont{Graesser}},
  \bibnamefont{and}
  \bibinfo{author}{\bibfnamefont{E.}~\bibnamefont{Mereghetti}}
  (\bibinfo{year}{2017}{\natexlab{a}}), \eprint{arXiv:1708.09390}.

\bibitem[{\citenamefont{Cirigliano
  et~al.}(2017{\natexlab{b}})\citenamefont{Cirigliano, Dekens, Graesser, and
  Mereghetti}}]{Cirigliano2017-PLB}
\bibinfo{author}{\bibfnamefont{V.}~\bibnamefont{Cirigliano}},
  \bibinfo{author}{\bibfnamefont{W.}~\bibnamefont{Dekens}},
  \bibinfo{author}{\bibfnamefont{M.}~\bibnamefont{Graesser}}, \bibnamefont{and}
  \bibinfo{author}{\bibfnamefont{E.}~\bibnamefont{Mereghetti}},
  \bibinfo{journal}{Physics Letters B} \textbf{\bibinfo{volume}{769}},
  \bibinfo{pages}{460 } (\bibinfo{year}{2017}{\natexlab{b}}), ISSN
  \bibinfo{issn}{0370-2693},
  \urlprefix\url{http://www.sciencedirect.com/science/article/pii/S0370269317302940}.

\bibitem[{\citenamefont{Berkowitz et~al.}(2017)\citenamefont{Berkowitz,
  Brantley, Bouchard, Chang, Clark, Garron, Joo, Kurth, Monahan, Monge-Camacho
  et~al.}}]{Loud2017}
\bibinfo{author}{\bibfnamefont{E.}~\bibnamefont{Berkowitz}},
  \bibinfo{author}{\bibfnamefont{D.}~\bibnamefont{Brantley}},
  \bibinfo{author}{\bibfnamefont{C.}~\bibnamefont{Bouchard}},
  \bibinfo{author}{\bibfnamefont{C.~C.} \bibnamefont{Chang}},
  \bibinfo{author}{\bibfnamefont{M.~A.} \bibnamefont{Clark}},
  \bibinfo{author}{\bibfnamefont{N.}~\bibnamefont{Garron}},
  \bibinfo{author}{\bibfnamefont{B.}~\bibnamefont{Joo}},
  \bibinfo{author}{\bibfnamefont{T.}~\bibnamefont{Kurth}},
  \bibinfo{author}{\bibfnamefont{C.}~\bibnamefont{Monahan}},
  \bibinfo{author}{\bibfnamefont{H.}~\bibnamefont{Monge-Camacho}},
  \bibnamefont{et~al.} (\bibinfo{year}{2017}), \eprint{arXiv:1704.01114}.

\bibitem[{\citenamefont{Hirsch et~al.}(1996{\natexlab{a}})\citenamefont{Hirsch,
  Klapdor-Kleingrothaus, and Kovalenko}}]{Hirsch1996plb}
\bibinfo{author}{\bibfnamefont{M.}~\bibnamefont{Hirsch}},
  \bibinfo{author}{\bibfnamefont{H.~V.} \bibnamefont{Klapdor-Kleingrothaus}},
  \bibnamefont{and} \bibinfo{author}{\bibfnamefont{S.~G.}
  \bibnamefont{Kovalenko}}, \bibinfo{journal}{Phys. Lett. B}
  \textbf{\bibinfo{volume}{372}}, \bibinfo{pages}{181}
  (\bibinfo{year}{1996}{\natexlab{a}}), \eprint{hep-ph/9512237}.

\bibitem[{\citenamefont{Pas et~al.}(1999)\citenamefont{Pas, Hirsch,
  Klapdor-Kleingrothaus, and Kovalenko}}]{Pas1999}
\bibinfo{author}{\bibfnamefont{H.}~\bibnamefont{Pas}},
  \bibinfo{author}{\bibfnamefont{M.}~\bibnamefont{Hirsch}},
  \bibinfo{author}{\bibfnamefont{H.~V.} \bibnamefont{Klapdor-Kleingrothaus}},
  \bibnamefont{and} \bibinfo{author}{\bibfnamefont{S.~G.}
  \bibnamefont{Kovalenko}}, \bibinfo{journal}{Phys. Lett. B}
  \textbf{\bibinfo{volume}{453}}, \bibinfo{pages}{194} (\bibinfo{year}{1999}).

\bibitem[{\citenamefont{Pas et~al.}(2001)\citenamefont{Pas, Hirsch,
  Klapdor-Kleingrothaus, and Kovalenko}}]{Pas2001}
\bibinfo{author}{\bibfnamefont{H.}~\bibnamefont{Pas}},
  \bibinfo{author}{\bibfnamefont{M.}~\bibnamefont{Hirsch}},
  \bibinfo{author}{\bibfnamefont{H.~V.} \bibnamefont{Klapdor-Kleingrothaus}},
  \bibnamefont{and} \bibinfo{author}{\bibfnamefont{S.~G.}
  \bibnamefont{Kovalenko}}, \bibinfo{journal}{Phys. Lett. B}
  \textbf{\bibinfo{volume}{498}}, \bibinfo{pages}{35} (\bibinfo{year}{2001}),
  \eprint{hep-ph/0008182}.

\bibitem[{\citenamefont{Deppisch et~al.}(2012)\citenamefont{Deppisch, Hirsch,
  and Pas}}]{Deppisch2012}
\bibinfo{author}{\bibfnamefont{F.~F.} \bibnamefont{Deppisch}},
  \bibinfo{author}{\bibfnamefont{M.}~\bibnamefont{Hirsch}}, \bibnamefont{and}
  \bibinfo{author}{\bibfnamefont{H.}~\bibnamefont{Pas}}, \bibinfo{journal}{J.
  Phys. G} \textbf{\bibinfo{volume}{39}}, \bibinfo{pages}{124007}
  (\bibinfo{year}{2012}).

\bibitem[{\citenamefont{Simkovic et~al.}(1999)\citenamefont{Simkovic, Pantis,
  Vergados, and Faessler}}]{Simkovic1999}
\bibinfo{author}{\bibfnamefont{F.}~\bibnamefont{Simkovic}},
  \bibinfo{author}{\bibfnamefont{G.}~\bibnamefont{Pantis}},
  \bibinfo{author}{\bibfnamefont{J.~D.} \bibnamefont{Vergados}},
  \bibnamefont{and} \bibinfo{author}{\bibfnamefont{A.}~\bibnamefont{Faessler}},
  \bibinfo{journal}{Phys. Rev. C} \textbf{\bibinfo{volume}{60}},
  \bibinfo{pages}{055502} (\bibinfo{year}{1999}).

\bibitem[{\citenamefont{Suhonen and Civitarese}(2010)}]{Suhonen2010}
\bibinfo{author}{\bibfnamefont{J.}~\bibnamefont{Suhonen}} \bibnamefont{and}
  \bibinfo{author}{\bibfnamefont{O.}~\bibnamefont{Civitarese}},
  \bibinfo{journal}{Nucl. Phys. A} \textbf{\bibinfo{volume}{847}},
  \bibinfo{pages}{207} (\bibinfo{year}{2010}).

\bibitem[{\citenamefont{Faessler et~al.}(2011)\citenamefont{Faessler, Meroni,
  Petcov, Simkovic, and Vergados}}]{Faessler2011}
\bibinfo{author}{\bibfnamefont{A.}~\bibnamefont{Faessler}},
  \bibinfo{author}{\bibfnamefont{A.}~\bibnamefont{Meroni}},
  \bibinfo{author}{\bibfnamefont{S.~T.} \bibnamefont{Petcov}},
  \bibinfo{author}{\bibfnamefont{F.}~\bibnamefont{Simkovic}}, \bibnamefont{and}
  \bibinfo{author}{\bibfnamefont{J.}~\bibnamefont{Vergados}},
  \bibinfo{journal}{Phys. Rev. D} \textbf{\bibinfo{volume}{83}},
  \bibinfo{pages}{113003} (\bibinfo{year}{2011}).

\bibitem[{\citenamefont{Mustonen and Engel}(2013)}]{MustonendEngel2013}
\bibinfo{author}{\bibfnamefont{M.~T.} \bibnamefont{Mustonen}} \bibnamefont{and}
  \bibinfo{author}{\bibfnamefont{J.}~\bibnamefont{Engel}},
  \bibinfo{journal}{Phys. Rev. C} \textbf{\bibinfo{volume}{87}},
  \bibinfo{pages}{064302} (\bibinfo{year}{2013}).

\bibitem[{\citenamefont{Faessler et~al.}(2014)\citenamefont{Faessler, Gonzalez,
  Kovalenko, and Simkovic}}]{FaesslerGonzales2014}
\bibinfo{author}{\bibfnamefont{A.}~\bibnamefont{Faessler}},
  \bibinfo{author}{\bibfnamefont{M.}~\bibnamefont{Gonzalez}},
  \bibinfo{author}{\bibfnamefont{S.}~\bibnamefont{Kovalenko}},
  \bibnamefont{and} \bibinfo{author}{\bibfnamefont{F.}~\bibnamefont{Simkovic}},
  \bibinfo{journal}{Phys. Rev. D} \textbf{\bibinfo{volume}{90}},
  \bibinfo{pages}{096010} (\bibinfo{year}{2014}).

\bibitem[{\citenamefont{Retamosa et~al.}(1995)\citenamefont{Retamosa, Caurier,
  and Nowacki}}]{Retamosa1995}
\bibinfo{author}{\bibfnamefont{J.}~\bibnamefont{Retamosa}},
  \bibinfo{author}{\bibfnamefont{E.}~\bibnamefont{Caurier}}, \bibnamefont{and}
  \bibinfo{author}{\bibfnamefont{F.}~\bibnamefont{Nowacki}},
  \bibinfo{journal}{Phys. Rev. C} \textbf{\bibinfo{volume}{51}},
  \bibinfo{pages}{371} (\bibinfo{year}{1995}).

\bibitem[{\citenamefont{Caurier et~al.}(1996)\citenamefont{Caurier, Nowacki,
  Poves, and Retamosa}}]{Caurier1996}
\bibinfo{author}{\bibfnamefont{E.}~\bibnamefont{Caurier}},
  \bibinfo{author}{\bibfnamefont{F.}~\bibnamefont{Nowacki}},
  \bibinfo{author}{\bibfnamefont{A.}~\bibnamefont{Poves}}, \bibnamefont{and}
  \bibinfo{author}{\bibfnamefont{J.}~\bibnamefont{Retamosa}},
  \bibinfo{journal}{Phys. Rev. Lett.} \textbf{\bibinfo{volume}{77}},
  \bibinfo{pages}{1954} (\bibinfo{year}{1996}).

\bibitem[{\citenamefont{Horoi}(2013)}]{Horoi2013}
\bibinfo{author}{\bibfnamefont{M.}~\bibnamefont{Horoi}},
  \bibinfo{journal}{Phys. Rev. C} \textbf{\bibinfo{volume}{87}},
  \bibinfo{pages}{014320} (\bibinfo{year}{2013}).

\bibitem[{\citenamefont{Neacsu and
  Stoica}(2014{\natexlab{a}})}]{Neacsu2014constraints}
\bibinfo{author}{\bibfnamefont{A.}~\bibnamefont{Neacsu}} \bibnamefont{and}
  \bibinfo{author}{\bibfnamefont{S.}~\bibnamefont{Stoica}},
  \bibinfo{journal}{Advances in High Energy Physics}
  \textbf{\bibinfo{volume}{2014}} (\bibinfo{year}{2014}{\natexlab{a}}).

\bibitem[{\citenamefont{Caurier et~al.}(2008)\citenamefont{Caurier, Menendez,
  Nowacki, and Poves}}]{Caurier2008}
\bibinfo{author}{\bibfnamefont{E.}~\bibnamefont{Caurier}},
  \bibinfo{author}{\bibfnamefont{J.}~\bibnamefont{Menendez}},
  \bibinfo{author}{\bibfnamefont{F.}~\bibnamefont{Nowacki}}, \bibnamefont{and}
  \bibinfo{author}{\bibfnamefont{A.}~\bibnamefont{Poves}},
  \bibinfo{journal}{Phys. Rev. Lett.} \textbf{\bibinfo{volume}{100}},
  \bibinfo{pages}{052503} (\bibinfo{year}{2008}).

\bibitem[{\citenamefont{Menendez et~al.}(2009)\citenamefont{Menendez, Poves,
  Caurier, and Nowacki}}]{MenendezPovesCaurier2009}
\bibinfo{author}{\bibfnamefont{J.}~\bibnamefont{Menendez}},
  \bibinfo{author}{\bibfnamefont{A.}~\bibnamefont{Poves}},
  \bibinfo{author}{\bibfnamefont{E.}~\bibnamefont{Caurier}}, \bibnamefont{and}
  \bibinfo{author}{\bibfnamefont{F.}~\bibnamefont{Nowacki}},
  \bibinfo{journal}{Nucl. Phys. A} \textbf{\bibinfo{volume}{818}},
  \bibinfo{pages}{139} (\bibinfo{year}{2009}).

\bibitem[{\citenamefont{Caurier et~al.}(2005)\citenamefont{Caurier,
  Martinez-Pinedo, Nowacki, Poves, and Zuker}}]{Caurier2005}
\bibinfo{author}{\bibfnamefont{E.}~\bibnamefont{Caurier}},
  \bibinfo{author}{\bibfnamefont{G.}~\bibnamefont{Martinez-Pinedo}},
  \bibinfo{author}{\bibfnamefont{F.}~\bibnamefont{Nowacki}},
  \bibinfo{author}{\bibfnamefont{A.}~\bibnamefont{Poves}}, \bibnamefont{and}
  \bibinfo{author}{\bibfnamefont{A.~P.} \bibnamefont{Zuker}},
  \bibinfo{journal}{Rev. Mod. Phys.} \textbf{\bibinfo{volume}{77}},
  \bibinfo{pages}{427} (\bibinfo{year}{2005}).

\bibitem[{\citenamefont{Horoi and Stoica}(2010)}]{HoroiStoica2010}
\bibinfo{author}{\bibfnamefont{M.}~\bibnamefont{Horoi}} \bibnamefont{and}
  \bibinfo{author}{\bibfnamefont{S.}~\bibnamefont{Stoica}},
  \bibinfo{journal}{Phys. Rev. C} \textbf{\bibinfo{volume}{81}},
  \bibinfo{pages}{024321} (\bibinfo{year}{2010}).

\bibitem[{\citenamefont{Neacsu et~al.}(2012)\citenamefont{Neacsu, Stoica, and
  Horoi}}]{NeacsuStoicaHoroi2012}
\bibinfo{author}{\bibfnamefont{A.}~\bibnamefont{Neacsu}},
  \bibinfo{author}{\bibfnamefont{S.}~\bibnamefont{Stoica}}, \bibnamefont{and}
  \bibinfo{author}{\bibfnamefont{M.}~\bibnamefont{Horoi}},
  \bibinfo{journal}{Phys. Rev. C} \textbf{\bibinfo{volume}{86}},
  \bibinfo{pages}{067304} (\bibinfo{year}{2012}).

\bibitem[{\citenamefont{Sen'kov and Horoi}(2013)}]{SenkovHoroi2013}
\bibinfo{author}{\bibfnamefont{R.~A.} \bibnamefont{Sen'kov}} \bibnamefont{and}
  \bibinfo{author}{\bibfnamefont{M.}~\bibnamefont{Horoi}},
  \bibinfo{journal}{Phys. Rev. C} \textbf{\bibinfo{volume}{88}},
  \bibinfo{pages}{064312} (\bibinfo{year}{2013}).

\bibitem[{\citenamefont{Horoi and Brown}(2013)}]{HoroiBrown2013}
\bibinfo{author}{\bibfnamefont{M.}~\bibnamefont{Horoi}} \bibnamefont{and}
  \bibinfo{author}{\bibfnamefont{B.~A.} \bibnamefont{Brown}},
  \bibinfo{journal}{Phys. Rev. Lett.} \textbf{\bibinfo{volume}{110}},
  \bibinfo{pages}{222502} (\bibinfo{year}{2013}).

\bibitem[{\citenamefont{Sen'kov et~al.}(2014)\citenamefont{Sen'kov, Horoi, and
  Brown}}]{SenkovHoroiBrown2014}
\bibinfo{author}{\bibfnamefont{R.~A.} \bibnamefont{Sen'kov}},
  \bibinfo{author}{\bibfnamefont{M.}~\bibnamefont{Horoi}}, \bibnamefont{and}
  \bibinfo{author}{\bibfnamefont{B.~A.} \bibnamefont{Brown}},
  \bibinfo{journal}{Phys. Rev. C} \textbf{\bibinfo{volume}{89}},
  \bibinfo{pages}{054304} (\bibinfo{year}{2014}).

\bibitem[{\citenamefont{Brown et~al.}(2014)\citenamefont{Brown, Horoi, and
  Sen'kov}}]{BrownHoroiSenkov2014}
\bibinfo{author}{\bibfnamefont{B.~A.} \bibnamefont{Brown}},
  \bibinfo{author}{\bibfnamefont{M.}~\bibnamefont{Horoi}}, \bibnamefont{and}
  \bibinfo{author}{\bibfnamefont{R.~A.} \bibnamefont{Sen'kov}},
  \bibinfo{journal}{Phys. Rev. Lett.} \textbf{\bibinfo{volume}{113}},
  \bibinfo{pages}{262501} (\bibinfo{year}{2014}).

\bibitem[{\citenamefont{Neacsu and
  Stoica}(2014{\natexlab{b}})}]{NeacsuStoica2014}
\bibinfo{author}{\bibfnamefont{A.}~\bibnamefont{Neacsu}} \bibnamefont{and}
  \bibinfo{author}{\bibfnamefont{S.}~\bibnamefont{Stoica}},
  \bibinfo{journal}{J. Phys. G} \textbf{\bibinfo{volume}{41}},
  \bibinfo{pages}{015201} (\bibinfo{year}{2014}{\natexlab{b}}).

\bibitem[{\citenamefont{Sen'kov and Horoi}(2014)}]{SenkovHoroi2014}
\bibinfo{author}{\bibfnamefont{R.~A.} \bibnamefont{Sen'kov}} \bibnamefont{and}
  \bibinfo{author}{\bibfnamefont{M.}~\bibnamefont{Horoi}},
  \bibinfo{journal}{Phys. Rev. C} \textbf{\bibinfo{volume}{90}},
  \bibinfo{pages}{{051301(R)}} (\bibinfo{year}{2014}).

\bibitem[{\citenamefont{Neacsu and Horoi}(2015)}]{NeacsuHoroi2015}
\bibinfo{author}{\bibfnamefont{A.}~\bibnamefont{Neacsu}} \bibnamefont{and}
  \bibinfo{author}{\bibfnamefont{M.}~\bibnamefont{Horoi}},
  \bibinfo{journal}{Phys. Rev. C} \textbf{\bibinfo{volume}{91}},
  \bibinfo{pages}{024309} (\bibinfo{year}{2015}).

\bibitem[{\citenamefont{Horoi and
  Neacsu}(2016{\natexlab{b}})}]{NeacsuHoroi2016}
\bibinfo{author}{\bibfnamefont{M.}~\bibnamefont{Horoi}} \bibnamefont{and}
  \bibinfo{author}{\bibfnamefont{A.}~\bibnamefont{Neacsu}},
  \bibinfo{journal}{Phys. Rev. C} \textbf{\bibinfo{volume}{93}},
  \bibinfo{pages}{024308} (\bibinfo{year}{2016}{\natexlab{b}}).

\bibitem[{\citenamefont{Horoi et~al.}(2007)\citenamefont{Horoi, Stoica, and
  Brown}}]{HoroiStoicaBrown2007}
\bibinfo{author}{\bibfnamefont{M.}~\bibnamefont{Horoi}},
  \bibinfo{author}{\bibfnamefont{S.}~\bibnamefont{Stoica}}, \bibnamefont{and}
  \bibinfo{author}{\bibfnamefont{B.~A.} \bibnamefont{Brown}},
  \bibinfo{journal}{Phys. Rev. C} \textbf{\bibinfo{volume}{75}},
  \bibinfo{pages}{034303} (\bibinfo{year}{2007}).

\bibitem[{\citenamefont{Blennow et~al.}(2010)\citenamefont{Blennow,
  Fernandez-Martinez, Lopez-Pavon, and Menendez}}]{Blennow2010}
\bibinfo{author}{\bibfnamefont{M.}~\bibnamefont{Blennow}},
  \bibinfo{author}{\bibfnamefont{E.}~\bibnamefont{Fernandez-Martinez}},
  \bibinfo{author}{\bibfnamefont{J.}~\bibnamefont{Lopez-Pavon}},
  \bibnamefont{and} \bibinfo{author}{\bibfnamefont{J.}~\bibnamefont{Menendez}},
  \bibinfo{journal}{JHEP} \textbf{\bibinfo{volume}{07}}, \bibinfo{pages}{096}
  (\bibinfo{year}{2010}).

\bibitem[{\citenamefont{Barea and Iachello}(2009)}]{Barea2009}
\bibinfo{author}{\bibfnamefont{J.}~\bibnamefont{Barea}} \bibnamefont{and}
  \bibinfo{author}{\bibfnamefont{F.}~\bibnamefont{Iachello}},
  \bibinfo{journal}{Phys. Rev. C} \textbf{\bibinfo{volume}{79}},
  \bibinfo{pages}{044301} (\bibinfo{year}{2009}).

\bibitem[{\citenamefont{Barea et~al.}(2012)\citenamefont{Barea, Kotila, and
  Iachello}}]{Barea2012}
\bibinfo{author}{\bibfnamefont{J.}~\bibnamefont{Barea}},
  \bibinfo{author}{\bibfnamefont{J.}~\bibnamefont{Kotila}}, \bibnamefont{and}
  \bibinfo{author}{\bibfnamefont{F.}~\bibnamefont{Iachello}},
  \bibinfo{journal}{Phys. Rev. Lett.} \textbf{\bibinfo{volume}{109}},
  \bibinfo{pages}{042501} (\bibinfo{year}{2012}).

\bibitem[{\citenamefont{Barea et~al.}(2013)\citenamefont{Barea, Kotila, and
  Iachello}}]{Barea2013}
\bibinfo{author}{\bibfnamefont{J.}~\bibnamefont{Barea}},
  \bibinfo{author}{\bibfnamefont{J.}~\bibnamefont{Kotila}}, \bibnamefont{and}
  \bibinfo{author}{\bibfnamefont{F.}~\bibnamefont{Iachello}},
  \bibinfo{journal}{Phys. Rev. C} \textbf{\bibinfo{volume}{87}},
  \bibinfo{pages}{014315} (\bibinfo{year}{2013}).

\bibitem[{\citenamefont{Barea et~al.}(2015)\citenamefont{Barea, Kotila, and
  Iachello}}]{Barea2015}
\bibinfo{author}{\bibfnamefont{J.}~\bibnamefont{Barea}},
  \bibinfo{author}{\bibfnamefont{J.}~\bibnamefont{Kotila}}, \bibnamefont{and}
  \bibinfo{author}{\bibfnamefont{F.}~\bibnamefont{Iachello}},
  \bibinfo{journal}{Phys. Rev. C} \textbf{\bibinfo{volume}{91}},
  \bibinfo{pages}{034304} (\bibinfo{year}{2015}).

\bibitem[{\citenamefont{Rath et~al.}(2013)\citenamefont{Rath, Chandra,
  Chaturvedi, Lohani, Raina, and Hirsch}}]{Rath2013}
\bibinfo{author}{\bibfnamefont{P.~K.} \bibnamefont{Rath}},
  \bibinfo{author}{\bibfnamefont{R.}~\bibnamefont{Chandra}},
  \bibinfo{author}{\bibfnamefont{K.}~\bibnamefont{Chaturvedi}},
  \bibinfo{author}{\bibfnamefont{P.}~\bibnamefont{Lohani}},
  \bibinfo{author}{\bibfnamefont{P.~K.} \bibnamefont{Raina}}, \bibnamefont{and}
  \bibinfo{author}{\bibfnamefont{J.~G.} \bibnamefont{Hirsch}},
  \bibinfo{journal}{Phys. Rev. C} \textbf{\bibinfo{volume}{88}},
  \bibinfo{pages}{064322} (\bibinfo{year}{2013}).

\bibitem[{\citenamefont{Rodriguez and Martinez-Pinedo}(2010)}]{Rodriguez2010}
\bibinfo{author}{\bibfnamefont{T.~R.} \bibnamefont{Rodriguez}}
  \bibnamefont{and}
  \bibinfo{author}{\bibfnamefont{G.}~\bibnamefont{Martinez-Pinedo}},
  \bibinfo{journal}{Phys. Rev. Lett.} \textbf{\bibinfo{volume}{105}},
  \bibinfo{pages}{252503} (\bibinfo{year}{2010}).

\bibitem[{\citenamefont{Song et~al.}(2014)\citenamefont{Song, Yao, Ring, and
  Meng}}]{Song2014}
\bibinfo{author}{\bibfnamefont{L.~S.} \bibnamefont{Song}},
  \bibinfo{author}{\bibfnamefont{J.~M.} \bibnamefont{Yao}},
  \bibinfo{author}{\bibfnamefont{P.}~\bibnamefont{Ring}}, \bibnamefont{and}
  \bibinfo{author}{\bibfnamefont{J.}~\bibnamefont{Meng}},
  \bibinfo{journal}{Phys. Rev. C} \textbf{\bibinfo{volume}{90}},
  \bibinfo{pages}{054309} (\bibinfo{year}{2014}).

\bibitem[{\citenamefont{Faessler et~al.}(2012)\citenamefont{Faessler, Rodin,
  and Simkovic}}]{Faessler2012}
\bibinfo{author}{\bibfnamefont{A.}~\bibnamefont{Faessler}},
  \bibinfo{author}{\bibfnamefont{V.}~\bibnamefont{Rodin}}, \bibnamefont{and}
  \bibinfo{author}{\bibfnamefont{F.}~\bibnamefont{Simkovic}},
  \bibinfo{journal}{J. Phys. G} \textbf{\bibinfo{volume}{39}},
  \bibinfo{pages}{124006} (\bibinfo{year}{2012}).

\bibitem[{\citenamefont{Vogel}(2012)}]{Vogel2012}
\bibinfo{author}{\bibfnamefont{P.}~\bibnamefont{Vogel}}, \bibinfo{journal}{J.
  Phys. G} \textbf{\bibinfo{volume}{39}}, \bibinfo{pages}{124002}
  (\bibinfo{year}{2012}).

\bibitem[{\citenamefont{Simkovic et~al.}(2013)\citenamefont{Simkovic, Rodin,
  Faessler, and Vogel}}]{SimkovicRodin2013}
\bibinfo{author}{\bibfnamefont{F.}~\bibnamefont{Simkovic}},
  \bibinfo{author}{\bibfnamefont{V.}~\bibnamefont{Rodin}},
  \bibinfo{author}{\bibfnamefont{A.}~\bibnamefont{Faessler}}, \bibnamefont{and}
  \bibinfo{author}{\bibfnamefont{P.}~\bibnamefont{Vogel}},
  \bibinfo{journal}{Phys. Rev. C} \textbf{\bibinfo{volume}{87}},
  \bibinfo{pages}{045501} (\bibinfo{year}{2013}).

\bibitem[{\citenamefont{Hyvarinen and Suhonen}(2015)}]{Suhonen2015}
\bibinfo{author}{\bibfnamefont{J.}~\bibnamefont{Hyvarinen}} \bibnamefont{and}
  \bibinfo{author}{\bibfnamefont{J.}~\bibnamefont{Suhonen}},
  \bibinfo{journal}{Phys. Rev. C} \textbf{\bibinfo{volume}{91}},
  \bibinfo{pages}{024613} (\bibinfo{year}{2015}).

\bibitem[{\citenamefont{Holt and Engel}(2013)}]{HoltEngel2013}
\bibinfo{author}{\bibfnamefont{J.~D.} \bibnamefont{Holt}} \bibnamefont{and}
  \bibinfo{author}{\bibfnamefont{J.}~\bibnamefont{Engel}},
  \bibinfo{journal}{Phys. Rev. C} \textbf{\bibinfo{volume}{87}},
  \bibinfo{pages}{064315} (\bibinfo{year}{2013}).

\bibitem[{\citenamefont{Sen'kov and Horoi}(2016)}]{Senkov2016}
\bibinfo{author}{\bibfnamefont{R.~A.} \bibnamefont{Sen'kov}} \bibnamefont{and}
  \bibinfo{author}{\bibfnamefont{M.}~\bibnamefont{Horoi}},
  \bibinfo{journal}{Phys. Rev. C} \textbf{\bibinfo{volume}{93}},
  \bibinfo{pages}{044334} (\bibinfo{year}{2016}).

\bibitem[{\citenamefont{Brown et~al.}(2015)\citenamefont{Brown, Fang, and
  Horoi}}]{BrownFangHoroi2015}
\bibinfo{author}{\bibfnamefont{B.~A.} \bibnamefont{Brown}},
  \bibinfo{author}{\bibfnamefont{D.~L.} \bibnamefont{Fang}}, \bibnamefont{and}
  \bibinfo{author}{\bibfnamefont{M.}~\bibnamefont{Horoi}},
  \bibinfo{journal}{Phys. Rev. C} \textbf{\bibinfo{volume}{92}},
  \bibinfo{pages}{041301} (\bibinfo{year}{2015}).

\bibitem[{\citenamefont{Doi et~al.}(1985)\citenamefont{Doi, Kotani, and
  Takasugi}}]{Doi1985}
\bibinfo{author}{\bibfnamefont{M.}~\bibnamefont{Doi}},
  \bibinfo{author}{\bibfnamefont{T.}~\bibnamefont{Kotani}}, \bibnamefont{and}
  \bibinfo{author}{\bibfnamefont{E.}~\bibnamefont{Takasugi}},
  \bibinfo{journal}{Prog. Theor. Phys. Suppl.} \textbf{\bibinfo{volume}{83}},
  \bibinfo{pages}{1} (\bibinfo{year}{1985}).

\bibitem[{\citenamefont{Stoica and Mirea}(2013)}]{StoicaMirea2013}
\bibinfo{author}{\bibfnamefont{S.}~\bibnamefont{Stoica}} \bibnamefont{and}
  \bibinfo{author}{\bibfnamefont{M.}~\bibnamefont{Mirea}},
  \bibinfo{journal}{Phys. Rev. C} \textbf{\bibinfo{volume}{88}},
  \bibinfo{pages}{037303} (\bibinfo{year}{2013}).

\bibitem[{\citenamefont{Arnold et~al.}(2016)\citenamefont{Arnold, Augier,
  Bakalyarov, Baker, Barabash, Basharina-Freshville, Blondel, Blot, Bongrand,
  Brudanin et~al.}}]{Nemo3-2016}
\bibinfo{author}{\bibfnamefont{R.}~\bibnamefont{Arnold}},
  \bibinfo{author}{\bibfnamefont{C.}~\bibnamefont{Augier}},
  \bibinfo{author}{\bibfnamefont{A.~M.} \bibnamefont{Bakalyarov}},
  \bibinfo{author}{\bibfnamefont{J.~D.} \bibnamefont{Baker}},
  \bibinfo{author}{\bibfnamefont{A.~S.} \bibnamefont{Barabash}},
  \bibinfo{author}{\bibfnamefont{A.}~\bibnamefont{Basharina-Freshville}},
  \bibinfo{author}{\bibfnamefont{S.}~\bibnamefont{Blondel}},
  \bibinfo{author}{\bibfnamefont{S.}~\bibnamefont{Blot}},
  \bibinfo{author}{\bibfnamefont{M.}~\bibnamefont{Bongrand}},
  \bibinfo{author}{\bibfnamefont{V.}~\bibnamefont{Brudanin}},
  \bibnamefont{et~al.} (\bibinfo{collaboration}{NEMO-3 Collaboration}),
  \bibinfo{journal}{Phys. Rev. D} \textbf{\bibinfo{volume}{93}},
  \bibinfo{pages}{112008} (\bibinfo{year}{2016}).

\bibitem[{\citenamefont{Agostini et~al.}(2017)\citenamefont{Agostini, Allardt,
  Bakalyarov, Balata, Barabanov, Baudis, Bauer, Bellotti, Belogurov, Belyaev
  et~al.}}]{Gerda-2017}
\bibinfo{author}{\bibfnamefont{M.}~\bibnamefont{Agostini}},
  \bibinfo{author}{\bibfnamefont{M.}~\bibnamefont{Allardt}},
  \bibinfo{author}{\bibfnamefont{A.}~\bibnamefont{Bakalyarov}},
  \bibinfo{author}{\bibfnamefont{M.}~\bibnamefont{Balata}},
  \bibinfo{author}{\bibfnamefont{I.}~\bibnamefont{Barabanov}},
  \bibinfo{author}{\bibfnamefont{L.}~\bibnamefont{Baudis}},
  \bibinfo{author}{\bibfnamefont{C.}~\bibnamefont{Bauer}},
  \bibinfo{author}{\bibfnamefont{E.}~\bibnamefont{Bellotti}},
  \bibinfo{author}{\bibfnamefont{S.}~\bibnamefont{Belogurov}},
  \bibinfo{author}{\bibfnamefont{S.}~\bibnamefont{Belyaev}},
  \bibnamefont{et~al.} (\bibinfo{collaboration}{GERDA Collaboration}),
  \bibinfo{journal}{Nature} \textbf{\bibinfo{volume}{544}}, \bibinfo{pages}{47}
  (\bibinfo{year}{2017}).

\bibitem[{Wat(2016)}]{Waters-2016}
\emph{\bibinfo{title}{{Latest results from NEMO-3 and status of the SuperNEMO
  Experiment}}} (\bibinfo{year}{2016}),
  \bibinfo{note}{\url{http://neutrino2016.iopconfs.org/IOP/media/uploaded/EVIOP/event_948/10.25__5__waters.pdf}}.

\bibitem[{\citenamefont{Alfonso et~al.}(2015)\citenamefont{Alfonso, Artusa,
  Avignone, Azzolini, Balata, Banks, Bari, Beeman, Bellini, Bersani
  et~al.}}]{Cuore-2015}
\bibinfo{author}{\bibfnamefont{K.}~\bibnamefont{Alfonso}},
  \bibinfo{author}{\bibfnamefont{D.~R.} \bibnamefont{Artusa}},
  \bibinfo{author}{\bibfnamefont{F.~T.} \bibnamefont{Avignone}},
  \bibinfo{author}{\bibfnamefont{O.}~\bibnamefont{Azzolini}},
  \bibinfo{author}{\bibfnamefont{M.}~\bibnamefont{Balata}},
  \bibinfo{author}{\bibfnamefont{T.~I.} \bibnamefont{Banks}},
  \bibinfo{author}{\bibfnamefont{G.}~\bibnamefont{Bari}},
  \bibinfo{author}{\bibfnamefont{J.~W.} \bibnamefont{Beeman}},
  \bibinfo{author}{\bibfnamefont{F.}~\bibnamefont{Bellini}},
  \bibinfo{author}{\bibfnamefont{A.}~\bibnamefont{Bersani}},
  \bibnamefont{et~al.} (\bibinfo{collaboration}{CUORE Collaboration}),
  \bibinfo{journal}{Phys. Rev. Lett.} \textbf{\bibinfo{volume}{115}},
  \bibinfo{pages}{102502} (\bibinfo{year}{2015}),
  \urlprefix\url{http://link.aps.org/doi/10.1103/PhysRevLett.115.102502}.

\bibitem[{\citenamefont{Gando et~al.}(2016)\citenamefont{Gando, Gando, Hachiya,
  Hayashi, Hayashida, Ikeda, Inoue, Ishidoshiro, Karino, Koga
  et~al.}}]{kamlandzen16}
\bibinfo{author}{\bibfnamefont{A.}~\bibnamefont{Gando}},
  \bibinfo{author}{\bibfnamefont{Y.}~\bibnamefont{Gando}},
  \bibinfo{author}{\bibfnamefont{T.}~\bibnamefont{Hachiya}},
  \bibinfo{author}{\bibfnamefont{A.}~\bibnamefont{Hayashi}},
  \bibinfo{author}{\bibfnamefont{S.}~\bibnamefont{Hayashida}},
  \bibinfo{author}{\bibfnamefont{H.}~\bibnamefont{Ikeda}},
  \bibinfo{author}{\bibfnamefont{K.}~\bibnamefont{Inoue}},
  \bibinfo{author}{\bibfnamefont{K.}~\bibnamefont{Ishidoshiro}},
  \bibinfo{author}{\bibfnamefont{Y.}~\bibnamefont{Karino}},
  \bibinfo{author}{\bibfnamefont{M.}~\bibnamefont{Koga}}, \bibnamefont{et~al.}
  (\bibinfo{collaboration}{KamLAND-Zen Collaboration}), \bibinfo{journal}{Phys.
  Rev. Lett.} \textbf{\bibinfo{volume}{117}}, \bibinfo{pages}{082503}
  (\bibinfo{year}{2016}),
  \urlprefix\url{https://link.aps.org/doi/10.1103/PhysRevLett.117.082503}.

\bibitem[{\citenamefont{Doi et~al.}(1983)\citenamefont{Doi, Kotani, Nishiura,
  and Takasugi}}]{Doi1983}
\bibinfo{author}{\bibfnamefont{M.}~\bibnamefont{Doi}},
  \bibinfo{author}{\bibfnamefont{T.}~\bibnamefont{Kotani}},
  \bibinfo{author}{\bibfnamefont{H.}~\bibnamefont{Nishiura}}, \bibnamefont{and}
  \bibinfo{author}{\bibfnamefont{E.}~\bibnamefont{Takasugi}},
  \bibinfo{journal}{Progr. Theor. Exp. Phys.} \textbf{\bibinfo{volume}{69}},
  \bibinfo{pages}{602} (\bibinfo{year}{1983}).

\bibitem[{\citenamefont{Rodejohann}(2012)}]{Rodejohann2012}
\bibinfo{author}{\bibfnamefont{W.}~\bibnamefont{Rodejohann}},
  \bibinfo{journal}{J. Phys. G} \textbf{\bibinfo{volume}{39}},
  \bibinfo{pages}{124008} (\bibinfo{year}{2012}).

\bibitem[{\citenamefont{Mohapatra and
  Pati}(1975{\natexlab{b}})}]{Mohapatra1975}
\bibinfo{author}{\bibfnamefont{R.~N.} \bibnamefont{Mohapatra}}
  \bibnamefont{and} \bibinfo{author}{\bibfnamefont{J.~C.} \bibnamefont{Pati}},
  \bibinfo{journal}{Phys. Rev. D} \textbf{\bibinfo{volume}{11}},
  \bibinfo{pages}{566} (\bibinfo{year}{1975}{\natexlab{b}}).

\bibitem[{\citenamefont{Hirsch et~al.}(1996{\natexlab{b}})\citenamefont{Hirsch,
  KlapdorKleingrothaus, and Kovalenko}}]{Hirsch1996}
\bibinfo{author}{\bibfnamefont{M.}~\bibnamefont{Hirsch}},
  \bibinfo{author}{\bibfnamefont{H.}~\bibnamefont{KlapdorKleingrothaus}},
  \bibnamefont{and}
  \bibinfo{author}{\bibfnamefont{S.}~\bibnamefont{Kovalenko}},
  \bibinfo{journal}{Phys. Rev. D} \textbf{\bibinfo{volume}{53}},
  \bibinfo{pages}{1329} (\bibinfo{year}{1996}{\natexlab{b}}).

\bibitem[{\citenamefont{Kolb et~al.}(1997)\citenamefont{Kolb, Hirsch, and
  Klapdor-Kleingrothaus}}]{Hirsch1997}
\bibinfo{author}{\bibfnamefont{S.}~\bibnamefont{Kolb}},
  \bibinfo{author}{\bibfnamefont{M.}~\bibnamefont{Hirsch}}, \bibnamefont{and}
  \bibinfo{author}{\bibfnamefont{H.~V.} \bibnamefont{Klapdor-Kleingrothaus}},
  \bibinfo{journal}{Phys. Rev. D} \textbf{\bibinfo{volume}{56}},
  \bibinfo{pages}{4161} (\bibinfo{year}{1997}).

\bibitem[{\citenamefont{Faessler et~al.}(2008)\citenamefont{Faessler, Gutsche,
  Kovalenko, and \ifmmode~\check{S}\else \v{S}\fi{}imkovic}}]{Kovalenko2008}
\bibinfo{author}{\bibfnamefont{A.}~\bibnamefont{Faessler}},
  \bibinfo{author}{\bibfnamefont{T.}~\bibnamefont{Gutsche}},
  \bibinfo{author}{\bibfnamefont{S.}~\bibnamefont{Kovalenko}},
  \bibnamefont{and}
  \bibinfo{author}{\bibfnamefont{F.}~\bibnamefont{\ifmmode~\check{S}\else
  \v{S}\fi{}imkovic}}, \bibinfo{journal}{Phys. Rev. D}
  \textbf{\bibinfo{volume}{77}}, \bibinfo{pages}{113012}
  (\bibinfo{year}{2008}).

\bibitem[{\citenamefont{Suhonen and Civitarese}(1998)}]{SuhonenCivitarese1998}
\bibinfo{author}{\bibfnamefont{J.}~\bibnamefont{Suhonen}} \bibnamefont{and}
  \bibinfo{author}{\bibfnamefont{O.}~\bibnamefont{Civitarese}},
  \bibinfo{journal}{Phys. Rep.} \textbf{\bibinfo{volume}{300}},
  \bibinfo{pages}{123} (\bibinfo{year}{1998}).

\bibitem[{\citenamefont{Kotila and Iachello}(2012)}]{Kotila2012}
\bibinfo{author}{\bibfnamefont{J.}~\bibnamefont{Kotila}} \bibnamefont{and}
  \bibinfo{author}{\bibfnamefont{F.}~\bibnamefont{Iachello}},
  \bibinfo{journal}{Phys. Rev. C} \textbf{\bibinfo{volume}{85}},
  \bibinfo{pages}{034316} (\bibinfo{year}{2012}).

\bibitem[{\citenamefont{Horoi and
  Neacsu}(2016{\natexlab{c}})}]{HoroiNeacsu2016psf}
\bibinfo{author}{\bibfnamefont{M.}~\bibnamefont{Horoi}} \bibnamefont{and}
  \bibinfo{author}{\bibfnamefont{A.}~\bibnamefont{Neacsu}},
  \bibinfo{journal}{Adv. High Energy Phys.} \textbf{\bibinfo{volume}{2016}},
  \bibinfo{pages}{7486712} (\bibinfo{year}{2016}{\natexlab{c}}).

\bibitem[{\citenamefont{Vergados et~al.}(2012)\citenamefont{Vergados, Ejiri,
  and Simkovic}}]{Vergados2012}
\bibinfo{author}{\bibfnamefont{J.~D.} \bibnamefont{Vergados}},
  \bibinfo{author}{\bibfnamefont{H.}~\bibnamefont{Ejiri}}, \bibnamefont{and}
  \bibinfo{author}{\bibfnamefont{F.}~\bibnamefont{Simkovic}},
  \bibinfo{journal}{Rep. Prog. Phys.} \textbf{\bibinfo{volume}{75}},
  \bibinfo{pages}{106301} (\bibinfo{year}{2012}).

\bibitem[{\citenamefont{Stefanik et~al.}(2015)\citenamefont{Stefanik,
  Dvornicky, Simkovic, and Vogel}}]{Stefanik2015}
\bibinfo{author}{\bibfnamefont{D.}~\bibnamefont{Stefanik}},
  \bibinfo{author}{\bibfnamefont{R.}~\bibnamefont{Dvornicky}},
  \bibinfo{author}{\bibfnamefont{F.}~\bibnamefont{Simkovic}}, \bibnamefont{and}
  \bibinfo{author}{\bibfnamefont{P.}~\bibnamefont{Vogel}},
  \bibinfo{journal}{Phys. Rev. C} \textbf{\bibinfo{volume}{92}},
  \bibinfo{pages}{055502} (\bibinfo{year}{2015}), \eprint{arXiv:1506.07145
  [hep-ph]}.

\bibitem[{\citenamefont{Muto et~al.}(1989)\citenamefont{Muto, Bender, and
  Klapdor}}]{Muto1989}
\bibinfo{author}{\bibfnamefont{K.}~\bibnamefont{Muto}},
  \bibinfo{author}{\bibfnamefont{E.}~\bibnamefont{Bender}}, \bibnamefont{and}
  \bibinfo{author}{\bibfnamefont{H.}~\bibnamefont{Klapdor}},
  \bibinfo{journal}{Z. Phys. A - Atomic Nuclei} \textbf{\bibinfo{volume}{334}},
  \bibinfo{pages}{187} (\bibinfo{year}{1989}).

\bibitem[{\citenamefont{Faessler and Simkovic}(1998)}]{Faessler1998-jpg}
\bibinfo{author}{\bibfnamefont{A.}~\bibnamefont{Faessler}} \bibnamefont{and}
  \bibinfo{author}{\bibfnamefont{F.}~\bibnamefont{Simkovic}},
  \bibinfo{journal}{Journal of Physics G: Nuclear and Particle Physics}
  \textbf{\bibinfo{volume}{24}}, \bibinfo{pages}{2139} (\bibinfo{year}{1998}),
  \urlprefix\url{http://stacks.iop.org/0954-3899/24/i=12/a=001}.

\bibitem[{\citenamefont{Wodecki et~al.}(1999)\citenamefont{Wodecki,
  Kami\ifmmode~\acute{n}\else \'{n}\fi{}ski, and \ifmmode~\check{S}\else
  \v{S}\fi{}imkovic}}]{Wodecki1999}
\bibinfo{author}{\bibfnamefont{A.}~\bibnamefont{Wodecki}},
  \bibinfo{author}{\bibfnamefont{W.~A.}
  \bibnamefont{Kami\ifmmode~\acute{n}\else \'{n}\fi{}ski}}, \bibnamefont{and}
  \bibinfo{author}{\bibfnamefont{F.}~\bibnamefont{\ifmmode~\check{S}\else
  \v{S}\fi{}imkovic}}, \bibinfo{journal}{Phys. Rev. D}
  \textbf{\bibinfo{volume}{60}}, \bibinfo{pages}{115007}
  (\bibinfo{year}{1999}).

\bibitem[{\citenamefont{Prezeau et~al.}(2003)\citenamefont{Prezeau,
  Ramsey-Musolf, and Vogel}}]{Prezeau2003}
\bibinfo{author}{\bibfnamefont{G.}~\bibnamefont{Prezeau}},
  \bibinfo{author}{\bibfnamefont{M.}~\bibnamefont{Ramsey-Musolf}},
  \bibnamefont{and} \bibinfo{author}{\bibfnamefont{P.}~\bibnamefont{Vogel}},
  \bibinfo{journal}{Phys.Rev. D} \textbf{\bibinfo{volume}{68}},
  \bibinfo{pages}{034016} (\bibinfo{year}{2003}).

\bibitem[{\citenamefont{Peng et~al.}(2016)\citenamefont{Peng, Ramsey-Musolf,
  and Winslow}}]{Peng2016}
\bibinfo{author}{\bibfnamefont{T.}~\bibnamefont{Peng}},
  \bibinfo{author}{\bibfnamefont{M.~J.} \bibnamefont{Ramsey-Musolf}},
  \bibnamefont{and} \bibinfo{author}{\bibfnamefont{P.}~\bibnamefont{Winslow}},
  \bibinfo{journal}{Phys. Rev. D} \textbf{\bibinfo{volume}{93}},
  \bibinfo{pages}{093002} (\bibinfo{year}{2016}).

\bibitem[{\citenamefont{Deppisch et~al.}(2015)\citenamefont{Deppisch, Harz,
  Huang, Hirsch, and P\"as}}]{Deppisch2015-PRD92}
\bibinfo{author}{\bibfnamefont{F.~F.} \bibnamefont{Deppisch}},
  \bibinfo{author}{\bibfnamefont{J.}~\bibnamefont{Harz}},
  \bibinfo{author}{\bibfnamefont{W.-C.} \bibnamefont{Huang}},
  \bibinfo{author}{\bibfnamefont{M.}~\bibnamefont{Hirsch}}, \bibnamefont{and}
  \bibinfo{author}{\bibfnamefont{H.}~\bibnamefont{P\"as}},
  \bibinfo{journal}{Phys. Rev. D} \textbf{\bibinfo{volume}{92}},
  \bibinfo{pages}{036005} (\bibinfo{year}{2015}).

\bibitem[{\citenamefont{Ahmed et~al.}(2017)\citenamefont{Ahmed, Neacsu, and
  Horoi}}]{Ahmed2017}
\bibinfo{author}{\bibfnamefont{F.}~\bibnamefont{Ahmed}},
  \bibinfo{author}{\bibfnamefont{A.}~\bibnamefont{Neacsu}}, \bibnamefont{and}
  \bibinfo{author}{\bibfnamefont{M.}~\bibnamefont{Horoi}},
  \bibinfo{journal}{Physics Letters B} \textbf{\bibinfo{volume}{769}},
  \bibinfo{pages}{299} (\bibinfo{year}{2017}).

\bibitem[{\citenamefont{Maki et~al.}(1962)\citenamefont{Maki, Nakagawa, and
  Sakata}}]{pmns}
\bibinfo{author}{\bibfnamefont{Z.}~\bibnamefont{Maki}},
  \bibinfo{author}{\bibfnamefont{M.}~\bibnamefont{Nakagawa}}, \bibnamefont{and}
  \bibinfo{author}{\bibfnamefont{S.}~\bibnamefont{Sakata}},
  \bibinfo{journal}{Prog. Theor. Phys.} \textbf{\bibinfo{volume}{28}},
  \bibinfo{pages}{870} (\bibinfo{year}{1962}),
  \urlprefix\url{http://dx.doi.org/10.1143/PTP.28.870}.

\bibitem[{\citenamefont{Beringer et~al.}(2012)\citenamefont{Beringer, Arguin,
  Barnett, Copic, Dahl, Groom, Lin, Lys, Murayama, Wohl et~al.}}]{upmns}
\bibinfo{author}{\bibfnamefont{J.}~\bibnamefont{Beringer}},
  \bibinfo{author}{\bibfnamefont{J.~F.} \bibnamefont{Arguin}},
  \bibinfo{author}{\bibfnamefont{R.~M.} \bibnamefont{Barnett}},
  \bibinfo{author}{\bibfnamefont{K.}~\bibnamefont{Copic}},
  \bibinfo{author}{\bibfnamefont{O.}~\bibnamefont{Dahl}},
  \bibinfo{author}{\bibfnamefont{D.~E.} \bibnamefont{Groom}},
  \bibinfo{author}{\bibfnamefont{C.~J.} \bibnamefont{Lin}},
  \bibinfo{author}{\bibfnamefont{J.}~\bibnamefont{Lys}},
  \bibinfo{author}{\bibfnamefont{H.}~\bibnamefont{Murayama}},
  \bibinfo{author}{\bibfnamefont{C.~G.} \bibnamefont{Wohl}},
  \bibnamefont{et~al.} (\bibinfo{collaboration}{Particle Data Group}),
  \bibinfo{journal}{Phys. Rev. D} \textbf{\bibinfo{volume}{86}},
  \bibinfo{pages}{010001} (\bibinfo{year}{2012}).

\bibitem[{\citenamefont{Adler et~al.}(1975)\citenamefont{Adler, Colglazier,
  Healy, Karliner, Lieberman, Ng, and Tsao}}]{Adler1975}
\bibinfo{author}{\bibfnamefont{S.~L.} \bibnamefont{Adler}},
  \bibinfo{author}{\bibfnamefont{E.~W.} \bibnamefont{Colglazier}},
  \bibinfo{author}{\bibfnamefont{J.~B.} \bibnamefont{Healy}},
  \bibinfo{author}{\bibfnamefont{I.}~\bibnamefont{Karliner}},
  \bibinfo{author}{\bibfnamefont{J.}~\bibnamefont{Lieberman}},
  \bibinfo{author}{\bibfnamefont{Y.~J.} \bibnamefont{Ng}}, \bibnamefont{and}
  \bibinfo{author}{\bibfnamefont{H.~S.} \bibnamefont{Tsao}},
  \bibinfo{journal}{Phys. Rev. D} \textbf{\bibinfo{volume}{11}},
  \bibinfo{pages}{3309} (\bibinfo{year}{1975}),
  \urlprefix\url{http://link.aps.org/doi/10.1103/PhysRevD.11.3309}.

\bibitem[{\citenamefont{Faessler et~al.}(1998)\citenamefont{Faessler,
  Kovalenko, and \ifmmode~\check{S}\else \v{S}\fi{}imkovic}}]{Faessler1998}
\bibinfo{author}{\bibfnamefont{A.}~\bibnamefont{Faessler}},
  \bibinfo{author}{\bibfnamefont{S.}~\bibnamefont{Kovalenko}},
  \bibnamefont{and}
  \bibinfo{author}{\bibfnamefont{F.}~\bibnamefont{\ifmmode~\check{S}\else
  \v{S}\fi{}imkovic}}, \bibinfo{journal}{Phys. Rev. D}
  \textbf{\bibinfo{volume}{58}}, \bibinfo{pages}{115004}
  (\bibinfo{year}{1998}).

\bibitem[{\citenamefont{Simkovic et~al.}(2009)\citenamefont{Simkovic, Faessler,
  Muether, Rodin, and Stauf}}]{Simkovic2009}
\bibinfo{author}{\bibfnamefont{F.}~\bibnamefont{Simkovic}},
  \bibinfo{author}{\bibfnamefont{A.}~\bibnamefont{Faessler}},
  \bibinfo{author}{\bibfnamefont{H.}~\bibnamefont{Muether}},
  \bibinfo{author}{\bibfnamefont{V.}~\bibnamefont{Rodin}}, \bibnamefont{and}
  \bibinfo{author}{\bibfnamefont{M.}~\bibnamefont{Stauf}},
  \bibinfo{journal}{Phys. Rev. C} \textbf{\bibinfo{volume}{79}},
  \bibinfo{pages}{055501} (\bibinfo{year}{2009}).

\bibitem[{\citenamefont{Honma et~al.}(2005)\citenamefont{Honma, Otsuka, Brown,
  and Mizusaki}}]{Honma2005}
\bibinfo{author}{\bibfnamefont{M.}~\bibnamefont{Honma}},
  \bibinfo{author}{\bibfnamefont{T.}~\bibnamefont{Otsuka}},
  \bibinfo{author}{\bibfnamefont{B.~A.} \bibnamefont{Brown}}, \bibnamefont{and}
  \bibinfo{author}{\bibfnamefont{T.}~\bibnamefont{Mizusaki}},
  \bibinfo{journal}{Eur. Phys. J. A} \textbf{\bibinfo{volume}{25 Suppl. 1}},
  \bibinfo{pages}{499} (\bibinfo{year}{2005}).

\bibitem[{\citenamefont{Honma et~al.}(2009)\citenamefont{Honma, Otsuka,
  Mizusaki, and Hjorth-Jensen}}]{JUN45}
\bibinfo{author}{\bibfnamefont{M.}~\bibnamefont{Honma}},
  \bibinfo{author}{\bibfnamefont{T.}~\bibnamefont{Otsuka}},
  \bibinfo{author}{\bibfnamefont{T.}~\bibnamefont{Mizusaki}}, \bibnamefont{and}
  \bibinfo{author}{\bibfnamefont{M.}~\bibnamefont{Hjorth-Jensen}},
  \bibinfo{journal}{Phys. Rev. C} \textbf{\bibinfo{volume}{80}},
  \bibinfo{pages}{064323} (\bibinfo{year}{2009}).

\bibitem[{\citenamefont{Qi and Xu}(2012)}]{Chong2012}
\bibinfo{author}{\bibfnamefont{C.}~\bibnamefont{Qi}} \bibnamefont{and}
  \bibinfo{author}{\bibfnamefont{Z.~X.} \bibnamefont{Xu}},
  \bibinfo{journal}{Phys. Rev. C} \textbf{\bibinfo{volume}{86}},
  \bibinfo{pages}{044323} (\bibinfo{year}{2012}).

\end{thebibliography}

\end{document}